\documentclass[prb,preprint,amsmath,amssymb,longbibliography]{revtex4-2}
\usepackage[utf8]{inputenc}
\usepackage{mathrsfs}
\usepackage{amsmath}
\usepackage{graphicx}



\begin{document}
\title{Stability and thermodynamic properties of bound magnetic polarons
in ferromagnetic semiconductors: Beyond the Gaussian approximation}
\author{Henryk Bednarski}
\affiliation{Centre of Polymer and Carbon Materials, Polish Academy of Sciences,
ul. M. Curie-Skłodowskiej 34, 41--819 Zabrze, Poland}
\date{\today}
\begin{abstract}
The Gaussian approximation for thermodynamic fluctuations of the host
magnetization, while successful for diluted magnetic semiconductors
far from magnetic ordering, fails for ferromagnetic semiconductors
such as GdN near their Curie temperature $T_{c}$. Diverging susceptibility
produces unphysical instabilities in the bound magnetic polaron (BMP)
free energy. Here we extend the Dietl--Spałek theory beyond the Gaussian
level by incorporating the full cubic and quartic anharmonic terms
of the Ginzburg--Landau--Wilson functional. Two complementary and
consistent generalizations are developed: (i) a perturbative non-local
treatment valid for finite spin correlation length $\xi$ and (ii)
a non-perturbative resummation of dominant local fluctuations into closed
exponential form in the strict local limit $\xi\to0$ inherent to
the original framework. The latter is rigorously justified by a novel
polaronic Ginzburg criterion, which shows that constraint-induced
non-localities are suppressed by $\mathcal{O}(1/N_{{\rm eff}})$ for
macroscopic polarons ($N_{{\rm eff}}\gg1$). Crucially, variational
optimization of the donor orbital radius is included, enabling a quantitative
description of magnetic self-trapping. When applied to GdN ($T_{c} \approx 55$ K), the theory eliminates Gaussian divergence and predicts thermodynamically stable, ferromagnetically ordered BMPs. These polarons exhibit a finite spontaneous internal spin splitting ($\bar{\Delta} \neq 0$) that persists deep into the paramagnetic phase. For a realistic exchange coupling of $J_{c} = 400$ meV—consistent with \textit{ab initio} estimates—this robust local ordering survives up to an effective temperature $T^{*} \approx 155$--$160$ K. The pronounced orbital contraction signals
magnetic self-trapping, with a characteristic kink in the temperature
dependence of the optimal Bohr radius at $T_{{\rm char}}\approx79$
K. Furthermore, the model reveals a critical exchange coupling threshold at $J_{c} \approx 330$ meV, above which the system undergoes a cooperative collapse into a highly localized, small polaron state. The model also identifies an optimal donor-concentration window
near the metal--insulator transition, where enhanced dielectric screening
maximizes $T^{*}$. These results establish a microscopic mechanism
for persistent BMP-mediated ferromagnetism well above the bulk $T_{c}$
via polaron percolation and suggest clear experimental signatures
(optical spin splitting, anomalous magnetoresistance, and susceptibility
tails) testable in GdN and related compounds (EuO, EuS, magnetic oxides). 
\end{abstract}
\maketitle

\section{Introduction}

The interplay between localized magnetic moments and shallow donor electrons in magnetic semiconductors leads to the formation of bound magnetic polarons (BMPs). Although introduced decades ago, this concept remains a highly active frontier in condensed matter physics. This enduring interest is driven by the fundamental role these polarons play in systems where the continuous spatial distribution of localized 3\textit{d} or 4\textit{f} moments strongly couples to shallow charge carriers. Recent experimental and theoretical studies continue to highlight the importance of BMPs in mediating bulk ferromagnetism, colossal magnetoresistance, and spintronic phenomena, particularly in ferromagnetic semiconductors and magnetic oxides \cite{Poh2025JMatChem,Alsmadi2025JAlloysComp,Kaur2021ApplSurfSci,Robkhob2020MaterSciEngB,Montalvo2026ACSOmega,Poornaprakash2021CeramInt,Alsmadi2023PRB,Bugajewski2025PRB}. For example, nitrogen-vacancy-mediated polarons have been shown to drive magnetic ordering in sputtered GdN thin films \cite{Bednarski2014NJP,Natali2013PRB}, while robust polaron percolation has been directly evidenced in Eu-based compounds such as EuCd${2}$P${2}$ \cite{Kopp2026npjQM}. Furthermore, the framework has been successfully extended to antiferromagnetic and altermagnetic systems \cite{Bugajewski2025PRB}. Direct spectroscopic observations of magnetic polaron states in materials like HgCr${2}$Se${4}$ \cite{Huang2026_2DMater}, alongside numerous demonstrations of BMP-driven ferromagnetism in 2D materials and dilute magnetic systems \cite{Vegesna2020SciRep,Kunj2020JIndEngChem,Islam2024MaterAdv,Sharma2023MaterResInnov,Maibam2020SNApplSci,Mrabet2022JAlloysComp,Siddique2020JAlloysComp,Ali2021JPhysChemC}, strongly underscore the ongoing need for refined theoretical models that properly capture non-Gaussian fluctuations near magnetic critical points.

A well-established theoretical framework for macroscopic BMPs, proposed
by Dietl--Spałek (DS) \cite{Dietl1983PRB}, successfully captures
the physics of these objects in diluted magnetic semiconductors (DMS).
The DS theory relies fundamentally on the Gaussian approximation for
the thermodynamic fluctuations of the localized spins. For DMS systems,
which are typically deep within the paramagnetic regime and lack a global
spontaneous magnetic ordering at finite temperatures, the Gaussian
model is entirely sufficient. The inclusion of Gaussian magnetization
fluctuations ($\boldsymbol{\eta}$) effectively smears out the macroscopic
magnetic instabilities predicted by simple mean-field approximations,
which adequately describes DMS systems deep in the paramagnetic phase.
However, a mathematical divergence emerges when applying this model
to ferromagnetic semiconductors such as GdN ($T_{c}=55$ K) or EuO.

In this context, we build upon prior extensions of the Dietl--Spałek
framework that accounted for interacting BMP pairs \cite{Bednarski2012JPCM,Bednarski2011APPA}
and incorporated the quartic term of the Ginzburg--Landau functional
to stabilize the mean-field magnetization \cite{Bednarski2014NJP}.
Although the resulting renormalized-Gaussian approach successfully
reproduced the double-dome carrier-concentration profile in GdN, it
treated magnetization fluctuations only up to quadratic (Gaussian)
order and kept the donor orbital size fixed at the bare hydrogenic
value. Consequently, it could neither capture the non-Gaussian character
of the fluctuation probability distribution nor describe the optimization
of the BMP orbital radius, leaving the physically important phenomenon
of magnetic self-trapping beyond its scope.

The present work overcomes these limitations by rigorously incorporating
the full cubic ($\propto M_{0}\eta_{\parallel}\boldsymbol{\eta}^{2}$)
and quartic ($\propto\boldsymbol{\eta}^{4}$) anharmonic terms of
the Ginzburg--Landau--Wilson functional directly into the polaron
partition function. Crucially, we perform a variational minimization
of the total free-energy functional with respect to the donor envelope
(hydrogenic trial function), while the most probable internal spin
splitting $\bar{\Delta}$ is extracted from the resulting probability
distribution. The stationary states are obtained from the renormalized
Schrödinger equation that follows from the Euler--Lagrange variation
of the total free-energy functional. This enables, for the first time
within a fully non-Gaussian framework, a complete microscopic description
of magnetic self-trapping --- the delicate competition between the
kinetic-energy cost of orbital contraction and the combined gain from
carrier--spin exchange and non-Gaussian (anharmonic) fluctuations
--- across the entire paramagnetic-to-ferromagnetic crossover.

Two complementary and fully consistent generalizations of the original
Dietl--Spałek theory are developed: (i) a non-local first-order perturbative
correction valid for arbitrary finite spin correlation length $\xi$,
and (ii) a non-perturbative resummation of the dominant local thermodynamic
fluctuations to closed exponential form in the strict local limit
$\xi\to0$ inherent to the DS framework. Going beyond the Gaussian
approximation is absolutely necessary near the ferromagnetic transition:
without the stabilizing effect of higher-order magnetization fluctuations,
the Gaussian model undergoes an unphysical divergence. In contrast, the
renormalized Gaussian model predicts anomalously low free energies
and unrealistic orbital collapse, whereas the fully nonlinear approach
leads to a stable, physically coherent polaron state.

The methodology of this paper is structured to provide both a general
perturbative perspective and a rigorous closed-form solution. We begin
by formulating the carrier-spin exchange in a continuous-medium model.
In Sec.~II.C we apply functional derivative techniques to the Ginzburg--Landau--Wilson
(GLW) functional \cite{ChaikinLubensky1995}. We first address the
general non-local case (finite spin correlation length $\xi$), where
a first-order perturbative expansion in $\lambda$ is employed. For
higher-order terms the non-locality introduces significant complexity,
so we subsequently transition to the strictly local limit $\xi\to0$---the
same local approximation that underlies the original Dietl--Spałek
theory \cite{Dietl1983PRB}. Within this limit we non-perturbatively
resum the leading local fluctuations into a closed exponential form,
neglecting sub-leading constraint-induced non-localities of order
$\mathcal{O}(1/N_{{\rm eff}})$, thereby providing a stable thermodynamic
description of BMPs across the Curie temperature.

When applied to the ferromagnetic semiconductor GdN, the non-perturbative
local theory eliminates the unphysical divergence inherent in both
the pure Gaussian and the renormalized-Gaussian treatments and yields
thermodynamically stable, ferromagnetically ordered bound magnetic
polarons carrying finite spontaneous spin splitting $\bar{\Delta}\neq0$
well into the paramagnetic phase $T>T_{c}$. Furthermore, our non-perturbative framework uncovers a counter-intuitive regime near the Mott metal-insulator transition, where the dielectric catastrophe robustly enhances the self-trapping of ferromagnetic droplets, offering a tuned window for high-temperature polaron percolation.

\section{Theoretical model}

\subsection{Carrier-spin exchange and the effective macroscopic splitting}

Consider a shallow donor electron interacting with a continuous
distribution of localized magnetic moments in a ferromagnetic semiconductor
(e.g., $4f$ spins of $Gd^{3+}$ ions). Assuming that the orbital
angular momentum is quenched and the carrier-spin exchange interaction
is moderately weak, the electronic envelope wavefunction $\varphi(\mathbf{r})$
can be effectively decoupled from the underlying spin degrees of freedom.
Following the continuous-medium approximation, the effective energy
of the donor electron can be expressed as the expectation value of
the single-particle Hamiltonian $\mathcal{H}_{1}$ modified by the
spin-splitting field $\boldsymbol{\Delta}[\mathbf{M};\varphi]$ \cite{Dietl1983PRB},
which was later adapted for ferromagnetic hosts \cite{Bednarski2014NJP}:
\begin{equation}
\mathcal{H}_{C}=\langle\varphi|\mathcal{H}_{1}|\varphi\rangle+s\boldsymbol{\gamma}\cdot\boldsymbol{\Delta}[\mathbf{M};\varphi].
\end{equation}

Here, $\boldsymbol{\gamma}$ represents the unit polarization vector
of the carrier spin (with eigenvalues $\pm1$), and $s$ is the carrier
spin quantum number ($s=1/2$ for the shallow donor electron). The
eigenvalues of the spin-dependent term are therefore $\pm s\Delta$,
and the local magnetization density of the host is given by $\mathbf{M}(\mathbf{r})$.
The effective spin-splitting field acting on the donor electron incorporates
both the external magnetic field $\mathbf{B}_{ext}$ and the continuous
$s-d$ ($s-f$) exchange interaction: 
\begin{equation}
\boldsymbol{\Delta}[\mathbf{M};\varphi]=\boldsymbol{\Delta}_{0}+\frac{\alpha}{g\mu_{B}}\int d^{3}r\left[\mathbf{M}(\mathbf{r})-\mathbf{M}_{0}\right]|\varphi(\mathbf{r})|^{2}.
\end{equation}

The vector $\mathbf{M}_{0}$ denotes the uniform, static magnetization
of the system, defining a natural quantization axis. Consequently,
the uniform part of the spin splitting is given by: 
\begin{equation}
\boldsymbol{\Delta}_{0}=g^{*}\mu_{B}\mathbf{B}_{ext}+\frac{\alpha}{g\mu_{B}}\mathbf{M}_{0},\label{eq:Delta0}
\end{equation}
where $g^{*}$ is the effective Landé factor of the electron, $\mu_{B}$
is the Bohr magneton, and $\alpha\sim\pm=J_{c}v_{0}/n_{0}$ is the
effective exchange coupling constant depending on the exchange integral
$J_{c}$ and the fractional volume per magnetic ion.

\subsection{The Ginzburg--Landau--Wilson functional and anharmonic fluctuations}

To capture the thermodynamics of the localized spins, particularly
in the vicinity of the ferromagnetic phase transition, we describe
the system using the generalized Ginzburg--Landau--Wilson (GLW)
free-energy functional. Crucially, to ensure thermodynamic stability
near the critical temperature $T_{c}$, we must retain the quartic
non-linearity. The functional is written as \cite{ZinnJustin2002}:
\begin{equation}
\mathcal{H}_{S}[\mathbf{M}]=\int\left\{ \frac{1}{2}\mu|\mathbf{M}(\mathbf{r})|^{2}+\frac{1}{2}\xi^{2}|\nabla\mathbf{M}(\mathbf{r})|^{2}+\frac{1}{4}\lambda|\mathbf{M}(\mathbf{r})|^{4}-\mathbf{H}_{a}\cdot\mathbf{M}(\mathbf{r})\right\} d^{3}r,\label{eq:GLW_functional}
\end{equation}
where $\mu$ is proportional to the inverse static magnetic susceptibility,
$\xi$ governs the correlation length, and $\lambda$ is the positive
phenomenological constant stabilizing the quartic term.

To analyze the thermodynamic fluctuations of the magnetization, we
separate the local magnetization field into its static mean-field
component and a spatially varying fluctuation field $\boldsymbol{\eta}(\mathbf{r})$:
\begin{equation}
\mathbf{M}(\mathbf{r})\equiv\mathbf{M}_{0}+\boldsymbol{\eta}(\mathbf{r})=\mathbf{M}_{0}+\boldsymbol{\eta}_{||}(\mathbf{r})+\boldsymbol{\eta}_{\perp}(\mathbf{r}),
\end{equation}
where $\boldsymbol{\eta}_{||}(\mathbf{r})$ and $\boldsymbol{\eta}_{\perp}(\mathbf{r})$
denote the longitudinal and transverse fluctuations with respect to
the $\mathbf{M}_{0}$ axis, respectively. Substituting this decomposition
into the GLW functional and expanding the terms $|\mathbf{M}(\mathbf{r})|^{2}$
and $|\mathbf{M}(\mathbf{r})|^{4}$, we can rigorously separate the
Hamiltonian into the macroscopic mean-field part $\mathcal{H}_{S}[\mathbf{M}_{0}]$,
the quadratic (Gaussian) fluctuation part $\mathcal{H}_{S}^{(2)}$,
and the higher-order anharmonic terms $\mathcal{H}_{S}^{(3,4)}$:
\begin{equation}
\mathcal{H}_{S}[\mathbf{M}_{0},\boldsymbol{\eta}]=\mathcal{H}_{S}[\mathbf{M}_{0}]+\mathcal{H}_{S}^{(2)}[\boldsymbol{\eta}]+\mathcal{H}_{S}^{(3,4)}[\boldsymbol{\eta}].
\end{equation}

The Gaussian part of the fluctuations absorbs the quadratic terms
stemming from the expansion of $\frac{1}{4}\lambda|\mathbf{M}(\mathbf{r})|^{4}$,
leading to renormalized, anisotropic inverse susceptibilities $\bar{\mu}_{||}$
and $\bar{\mu}_{\perp}$: 
\begin{equation}
\mathcal{H}_{S}^{(2)}[\boldsymbol{\eta}]=\sum_{i\,=\parallel,\perp}\int d^{3}r\left[\frac{1}{2}\bar{\mu}_{i}\,\eta_{i}^{2}(\mathbf{r})+\frac{1}{2}\xi^{2}|\nabla\eta_{i}(\mathbf{r})|^{2}\right],
\end{equation}
with the dressed parameters defined as $\bar{\mu}_{||}=\mu+3\lambda M_{0}^{2}$
and $\bar{\mu}_{\perp}=\mu+\lambda M_{0}^{2}$.

In the previous attempt to extend the BMP theory to ferromagnetic
semiconductors, carried out by Bednarski and Spałek \cite{Bednarski2014NJP},
only the contribution of the static magnetization $\mathbf{M}_{0}$
in the fourth power was taken into account. The magnetization fluctuations
$\boldsymbol{\eta}(\mathbf{r})$ were treated only up to the quadratic
(Gaussian) terms generated from the expansion of $|\mathbf{M}(\mathbf{r})|^{4}$,
while the higher-order anharmonic contributions $\mathcal{H}_{S}^{(3,4)}[\boldsymbol{\eta}]$
were neglected. Moreover, the donor orbital size was kept fixed at
the bare hydrogenic value $a_{B0}$. In ferromagnetic semiconductors
near the Curie temperature $T_{c}$, such a Gaussian-level truncation
of the fluctuation spectrum leads to unphysical instabilities and
divergences in the polaron free energy. At the same time, the lack
of variational optimization of the donor envelope precludes the description
of magnetic self-trapping --- the important physical effect arising
from the competition between kinetic energy and the gain from carrier--spin
exchange and non-Gaussian fluctuations. The current approach also
offers a more rigorous treatment of the auxiliary functional and the
resulting polaronic free energy compared to our previous preliminary
investigation \cite{Bednarski2014NJP}. Specifically, the definitions
of the convolution operators and the partition function normalization
have been refined here to ensure exact consistency with the original
Dietl--Spałek limit. Therefore, here we strictly preserve the non-linear
fluctuation functional: 
\begin{equation}
\mathcal{H}_{S}^{(3,4)}[\boldsymbol{\eta}]=\int\left\{ \lambda M_{0}\eta_{||}(\mathbf{r})|\boldsymbol{\eta}(\mathbf{r})|^{2}+\frac{1}{4}\lambda|\boldsymbol{\eta}(\mathbf{r})|^{4}\right\} d^{3}r.
\end{equation}

Incorporating these anharmonic terms into the total polaronic free
energy without introducing the divergence characteristic of unstable
perturbative expansions is the primary objective of the following
sections.

\subsection{Spin splitting distribution}

To obtain the renormalized (by fluctuations) Hamiltonian for the donor
electron, we define the probability distribution of the magnetization
fluctuations within the electron state $\varphi(\mathbf{r})$: 
\begin{equation}
\mathscr{P}[\boldsymbol{\eta};\mu,\lambda]\equiv\frac{\exp\left\{ -\beta\mathcal{H}_{S}[\boldsymbol{\eta};\mu,\lambda]\right\} }{\int D\boldsymbol{\eta}\exp\left\{ -\beta\mathcal{H}_{S}[\boldsymbol{\eta};\mu,\lambda]\right\} },\label{eq:prob-dist}
\end{equation}
where $\beta\equiv(k_{B}T)^{-1}$and $k_{B}$ is the Boltzmann constant.
The probability distribution of electron energies $P(E_{R})$ is then
\begin{align}
P(E_{R}) & =C\sum_{\gamma=\pm1}\int D\,\boldsymbol{\eta}\mathscr{P}[\boldsymbol{\eta};\mu,\lambda]\exp[-\beta(\mathcal{H}_{C})]\nonumber \\
 & =2C\exp\left[-\frac{\mathcal{H}_{1}}{k_{B}T}\right]\int D\boldsymbol{\eta}\cosh\left\{ s\beta\Delta[\boldsymbol{\eta}]\right\} \exp\left[-\frac{\mathcal{H}_{S}}{k_{B}T}\right],\label{eq:P-E-R}
\end{align}
with spin-only normalization $C^{-1}\equiv\int D\boldsymbol{\eta}\exp\{-\beta\mathcal{H}_{S}\}$.
Here $P[\boldsymbol{\eta}]$ is the pure-magnetization probability
(eq.~(\ref{eq:prob-dist})), and the carrier trace (the factor 2
cosh) is calculated out explicitly. The effective renormalized Hamiltonian
of the BMP can be defined as 
\begin{equation}
\mathcal{H}_{{\rm eff}}\equiv-k_{B}T\ln P(E_{R}).\label{eq:H-eff-def}
\end{equation}
A simpler and more convenient quantity is the probability distribution
$P(\boldsymbol{\Delta})$ of the spin splitting $\boldsymbol{\Delta}$:
\begin{equation}
P(\boldsymbol{\Delta})=\int\mathscr{P}[\boldsymbol{\eta}(\mathbf{r})]\exp\{-\beta\mathcal{H}_{C}\}\,\delta^{3}\{\boldsymbol{\Delta}-\boldsymbol{\Delta}[\boldsymbol{\eta}(\mathbf{r})]\}D\boldsymbol{\eta}(\mathbf{r}).\label{eq:P-Delta}
\end{equation}
The meaning of the transformation within the Dirac delta function
is explained in detail in the original DS work. Here we only note
that it bridges the macroscopic thermodynamics of the host lattice
with the polaronic state. Moreover, by substituting the macroscopic
magnetization $M_{0}$ with the uniform spin splitting parameter $\Delta_{0}$
via $M_{0}=\frac{g\mu_{B}}{\alpha}(\Delta_{0}-g^{*}\mu_{B}H_{a})$,
we can express the resulting effective probability distribution entirely
in terms of energy parameters. This approach maps the magnetic phase
transition directly onto the polaron's spin splitting, allowing us
to map the effective free-energy landscape (and its spontaneous symmetry
breaking) without relying on a loosely defined macroscopic polaron
volume near $T_{c}$. Performing the change of variables yields 
\begin{equation}
P(\boldsymbol{\Delta})=\mathcal{N}\exp\!\left(-\frac{\mathcal{H}_{1}}{k_{B}T}\right)\cosh\!\left(s\frac{\Delta}{k_{B}T}\right)Z[\boldsymbol{\Delta};\mathbf{J}(\mathbf{r})],\label{eq:P-Delta-2}
\end{equation}
where $\mathcal{N}$ is the single overall normalization constant chosen so
that $\int d^{3}\Delta\,P(\boldsymbol{\Delta})=1$ and the auxiliary
functional is 
\begin{align}
Z[\mathbf{J}] & =\int_{-\infty}^{\infty}D\boldsymbol{\eta}(\mathbf{r})\exp\Biggl(-\beta\int\Bigl\{\bigl[\bar{\mu}_{\parallel}\,\frac{\eta_{\parallel}^{2}(\mathbf{r})}{2}+\frac{\xi^{2}}{2}[\nabla\eta_{\parallel}(\mathbf{r})]^{2}\bigr]\nonumber \\
 & \quad+\bigl[\bar{\mu}_{\perp}\frac{\boldsymbol{\eta}_{\perp}^{2}(\mathbf{r})}{2}+\frac{\xi^{2}}{2}[\nabla\boldsymbol{\eta}_{\perp}(\mathbf{r})]^{2}\bigr]-\mathbf{J}(\mathbf{r})\cdot\boldsymbol{\eta}(\mathbf{r})\nonumber \\
 & \quad+\lambda M_{0}\eta_{\parallel}(\mathbf{r})\boldsymbol{\eta}^{2}(\mathbf{r})+\frac{1}{4}\lambda\boldsymbol{\eta}^{4}(\mathbf{r})\Bigr\} d^{3}r\Biggr)\delta^{3}\!\bigl\{\boldsymbol{\Delta}-\boldsymbol{\Delta}[\boldsymbol{\eta}(\mathbf{r})]\bigr\}.\label{eq:Z-J}
\end{align}
Here we have used the explicit expressions 
\begin{gather}
\boldsymbol{\eta}^{4}=(\boldsymbol{\eta}^{2})^{2}=\sum_{i}(\eta_{i}^{4}+2\sum_{j<i}\eta_{i}^{2}\eta_{j}^{2}),\label{eq:eta4}\\
\eta_{\parallel}\boldsymbol{\eta}^{2}=\eta_{3}^{3}+\eta_{3}(\eta_{1}^{2}+\eta_{2}^{2}).\label{eq:eta-par-eta2}
\end{gather}

The auxiliary functional $Z[\mathbf{J}]$ is evaluated using the generating-functional
technique \cite{ZinnJustin2002}.

From now on we include $\beta$ in all interaction parameters appearing
in $Z[\mathbf{J}]$ (i.e., $\bar{\mu}_{i}\to\tilde{\mu}_{i}=\beta\bar{\mu}_{i}$,
$\lambda\to\tilde{\lambda}=\beta\lambda$, the source $\mathbf{J}\to\tilde{\mathbf{J}}=\beta\mathbf{J}$,
and the gradient coefficient $\xi^{2}\to\tilde{\xi}^{2}=\beta\xi^{2}$).
This convention removes the explicit factor $\beta$ from the exponent,
greatly simplifying the combinatorial structure of the perturbative
expansion and the local resummation derived below. All expressions
for $\mathcal{F}_{i}$, $\mathcal{F}'_{i}$, $\Xi[\mathbf{J}]$, the
operator $\mathcal{O}$, and the propagator $Q_{i}$ are therefore
written with the rescaled (tilded) parameters. Note that this formal
absorption guarantees that the resulting propagator $Q_{i}(\mathbf{r},\mathbf{r}')$
(Eq.~(\ref{eq:propagator})) naturally functions as the thermal correlation
kernel. Consequently, the temperature dependence enters the diamond
convolution operator $\diamond_{i}$ and the variance $\Phi_{i}$
implicitly through the host fluctuations. We will restore the explicit
temperature dependence (and the original dimensions) when constructing
the final physical probability distribution $P(\boldsymbol{\Delta})$
and the free-energy functional in the subsequent sections.

To this end we introduce the diamond convolution operator 
\begin{equation}
A_{i}(\mathbf{r})\diamond_{i}B(\mathbf{r}')\equiv\frac{1}{2}\iint A_{i}(\mathbf{r})\,Q_{i}(\mathbf{r},\mathbf{r}')\,B(\mathbf{r}')\,d^{3}r\,d^{3}r',\label{eq:diamond}
\end{equation}
the propagator (Green function) of the quadratic GLW theory 
\begin{equation}
Q_{i}(\mathbf{r},\mathbf{r}')\equiv\int\frac{d^{3}k}{(2\pi)^{3}}\frac{e^{i\mathbf{k}\cdot(\mathbf{r}-\mathbf{r}')}}{\tilde{\mu}_{i}+k^{2}\tilde{\xi}^{2}}=\frac{1}{4\pi\tilde{\xi}^{2}|\mathbf{r}-\mathbf{r}'|}e^{-\frac{\sqrt{\tilde{\mu}_{i}}}{\tilde{\xi}}|\mathbf{r}-\mathbf{r}'|},\label{eq:propagator}
\end{equation}
as well as the quantities 
\begin{equation}
\Lambda_{i}\equiv\frac{\Delta_{i}-\Delta_{0i}}{2},\qquad\Phi_{i}\equiv|\varphi(\mathbf{r})|^{2}\diamond_{i}|\varphi(\mathbf{r}')|^{2}\Bigl(\frac{\alpha}{g\mu_{B}}\Bigr)^{2}.\label{eq:Lambda-Phi}
\end{equation}

The auxiliary functional then reads (with rescaled parameters) 
\begin{align}
Z[\tilde{\mathbf{J}}] & =\int_{-\infty}^{\infty}D\boldsymbol{\eta}(\mathbf{r})\exp\Biggl(-\int\Bigl\{\bigl[\tilde{\mu}_{\parallel}\,\frac{\eta_{\parallel}^{2}(\mathbf{r})}{2}+\frac{\tilde{\xi}^{2}}{2}[\nabla\eta_{\parallel}(\mathbf{r})]^{2}\bigr]\nonumber \\
 & \quad+\bigl[\tilde{\mu}_{\perp}\frac{\boldsymbol{\eta}_{\perp}^{2}(\mathbf{r})}{2}+\frac{\tilde{\xi}^{2}}{2}[\nabla\boldsymbol{\eta}_{\perp}(\mathbf{r})]^{2}\bigr]-\tilde{\mathbf{J}}(\mathbf{r})\cdot\boldsymbol{\eta}(\mathbf{r})\nonumber \\
 & \quad+\tilde{\lambda}M_{0}\eta_{\parallel}(\mathbf{r})\boldsymbol{\eta}^{2}(\mathbf{r})+\frac{1}{4}\tilde{\lambda}\boldsymbol{\eta}^{4}(\mathbf{r})\Bigr\} d^{3}r\Biggr)\delta^{3}\!\bigl\{\boldsymbol{\Delta}-\boldsymbol{\Delta}[\boldsymbol{\eta}(\mathbf{r})]\bigr\}.\label{eq:Z-J-tilded}
\end{align}
The Gaussian part of the functional (i.e., the unnormalized generating
functional for the quadratic GLW theory) then reads 
\begin{equation}
\Xi[\tilde{\mathbf{J}}]=\prod_{i=1}^{3}\exp\Bigl\{-\Bigl[\Phi_{i}^{-1}\bigl(\Lambda_{i}-\frac{\alpha}{g\mu_{B}}\tilde{J}_{i}(\mathbf{r})\diamond_{i}|\varphi(\mathbf{r}')|^{2}\bigr)^{2}-\tilde{J}_{i}(\mathbf{r})\diamond_{i}\tilde{J}_{i}(\mathbf{r}')\Bigr]\Bigr\},\label{eq:Xi-tilded}
\end{equation}
where the overall normalization constant of the full spin partition
function will be fixed at the end when constructing $P(\boldsymbol{\Delta})$
or $\Delta F$. It is instructive to clarify the derivation of Eq.~(\ref{eq:Xi-tilded})
and its exact relation to the original Dietl--Spałek framework \cite{Dietl1983PRB}.
While DS evaluated the constrained Gaussian integral by completing
the square to eliminate the linear magnetization term in the absence
of an external fluctuation source (cf. Eqs.~(C5) and (C6) in Appendix
C of Ref.~\cite{Dietl1983PRB}), maintaining the full generating
functional requires a more delicate, two-step procedure. First, the
completion of the square in the Ginzburg--Landau--Wilson quadratic
action to absorb the source $\tilde{\mathbf{J}}$ induces a non-local
shift in the fluctuation field, $\boldsymbol{\eta}\to\boldsymbol{\eta}'$.
Crucially, this shift propagates into the Dirac-delta constraint defining
the macroscopic spin splitting. As a result, the constraint itself
is displaced, generating the cross-term $\propto\tilde{\mathbf{J}}\diamond_{i}|\varphi|^{2}$.
A subsequent constrained minimization over the shifted variables yields
the exponential form. By rigorously absorbing the resulting field-independent
Gaussian integration constants (determinants) into the global normalization
factor $\mathcal{N}$ introduced in Eq.~(13), we arrive at the exact
equality presented in Eq.~(\ref{eq:Xi-tilded}). This explicit retention
of the source-dependent functional goes fundamentally beyond the original
DS treatment, providing the mathematical formalism necessary to extract
the higher-order non-Gaussian cumulants via functional derivatives.
Consequently, the auxiliary functions $\mathcal{F}_{i}(\mathbf{z})$
and $\mathcal{F}'_{i}(\mathbf{z})$ generated by these derivatives
(defined below) are not merely mathematical artifacts; they possess
a strict physical interpretation within the restricted Gaussian ensemble.
Specifically, $\mathcal{F}_{i}(\mathbf{z})\equiv\langle\eta_{i}(\mathbf{z})\rangle_{\tilde{\mathbf{J}}=0}$
represents the local linear response of the host magnetization to
the effective field of the donor electron, while $\mathcal{F}'_{i}(\mathbf{z})\equiv\langle\eta_{i}(\mathbf{z})^{2}\rangle_{c}$
is its local connected variance. Integrating these local fields over
the effective polaron volume connects our microscopic formulation
directly to the macroscopic thermodynamic quantities discussed by
DS. For instance, the volume-averaged first cumulant reproduces the
mean macroscopic magnetization relation $\overline{\Delta}\propto\langle M\rangle$
(cf. Eq.~(3.31) in Ref.~\cite{Dietl1983PRB}), and the global spatial
integral of the local variance scales as $2k_{B}T\chi/V_{p}$, exactly
recovering the macroscopic thermodynamic fluctuation limit $\langle M^{2}\rangle$
expressed in Eq.~(3.32) of the DS theory \cite{Dietl1983PRB}. (Note
that our auxiliary parameter $\Lambda_{i}$ corresponds directly to
the shift parameter $\lambda_{i}$ defined in Eq.~(C3) of Ref.~\cite{Dietl1983PRB},
but with the exchange coupling factor explicitly extracted to maintain
dimensional consistency with the source terms). The auxiliary functions
appearing in the anharmonic expansion are 
\begin{align}
\mathcal{F}[\tilde{J}_{i}(\mathbf{z})] & =\Bigl\{\Phi_{i}^{-1}[\Lambda_{i}-\frac{\alpha}{g\mu_{B}}\tilde{J}_{i}(\mathbf{r})\diamond_{i}|\varphi(\mathbf{r}')|^{2}]\left(\frac{\alpha}{g\mu_{B}}\right)\int Q_{i}(\mathbf{r},\mathbf{z})|\varphi(\mathbf{r})|^{2}\,d^{3}r\nonumber \\
 & \quad+\int\tilde{J}_{i}(\mathbf{r})Q_{i}(\mathbf{r},\mathbf{z})\,d^{3}r\Bigr\},\label{eq:F-J-tilded}\\
\mathcal{F}'[\tilde{J}_{i}(\mathbf{z})] & =\frac{\delta\mathcal{F}[\tilde{J}_{i}(\mathbf{z})]}{\delta\tilde{J}_{i}(\mathbf{z})}=-\frac{1}{2}\Phi_{i}^{-1}\Bigl(\frac{\alpha}{g\mu_{B}}\int Q_{i}(\mathbf{z},\mathbf{r}')|\varphi(\mathbf{r}')|^{2}\,d^{3}r'\Bigr)^{2}+Q_{i}(\mathbf{z},\mathbf{z}).\label{eq:Fprime-J-tilded}
\end{align}
The factor $\alpha/g\mu_{B}$ that appears explicitly in $\Xi[\tilde{\mathbf{J}}]$,
$\mathcal{F}$, and $\mathcal{F}'$ converts the magnetization-fluctuation
source $\tilde{\mathbf{J}}$ into the proper spin-splitting units
demanded by the Dirac-delta constraint. Setting now $\tilde{\mathbf{J}}(\mathbf{r})=\mathbf{0}$
we obtain the explicit building blocks 
\begin{equation}
\Xi[\tilde{\mathbf{J}}=0]=\exp\Bigl\{-\sum_{i}\Phi_{i}^{-1}\Lambda_{i}^{2}\Bigr\},\label{eq:Xi-zero-tilded}
\end{equation}
\begin{equation}
\mathcal{F}[\tilde{J}_{i}=0]=\Lambda_{i}\Phi_{i}^{-1}\left(\frac{\alpha}{g\mu_{B}}\right)\Psi_{i}(\mathbf{z}),\qquad\mathcal{F}'_{i}[\tilde{J}_{i}=0]=-\frac{1}{2}\Phi_{i}^{-1}\left(\frac{\alpha}{g\mu_{B}}\right)^{2}\Psi_{i}^{2}(\mathbf{z})+q_{i},\label{eq:F-Fprime-zero-tilded}
\end{equation}
where $\Psi_{i}(\mathbf{z})\equiv\int Q_{i}(\mathbf{r},\mathbf{z})|\varphi(\mathbf{r})|^{2}\,d^{3}r$
and $q_{i}\equiv Q_{i}(\mathbf{z},\mathbf{z})$ (this quantity will
be subjected to standard regularization later in the article). For
the physically relevant case of a shallow donor described by a hydrogenic
1s envelope, the convolutions $\Psi_{i}(\mathbf{z})$ and the parameters
$\Phi_{i}$ can be evaluated analytically; the explicit closed-form
derivations are presented in Appendix D.

The perturbative expansion of the auxiliary functional $Z[\tilde{\mathbf{J}}]$
in powers of the anharmonicity parameter $\tilde{\lambda}$ is obtained
by replacing the interaction terms $\tilde{\lambda}M_{0}\eta_{\parallel}(\mathbf{r})\boldsymbol{\eta}^{2}(\mathbf{r})+\frac{\tilde{\lambda}}{4}\boldsymbol{\eta}^{4}(\mathbf{r})$
with the corresponding functional derivatives acting on the Gaussian
generating functional $\Xi[\tilde{\mathbf{J}}]$. This yields the
formally exact power series 
\begin{align}
Z[\tilde{\mathbf{J}}] & =\sum_{n=0}^{\infty}\frac{1}{n!}\left\{ -\tilde{\lambda}\int d^{3}z\left[M_{0}\left(\frac{\delta^{3}}{\delta\tilde{J}_{3}^{3}(z)}+\frac{\delta^{3}}{\delta\tilde{J}_{3}(z)\delta\tilde{J}_{1}^{2}(z)}+\frac{\delta^{3}}{\delta\tilde{J}_{3}(z)\delta\tilde{J}_{2}^{2}(z)}\right)\right.\right.\nonumber \\
 & \quad\left.\left.+\frac{1}{4}\left(\sum_{i=1}^{3}\frac{\delta^{4}}{\delta\tilde{J}_{i}^{4}(z)}+2\sum_{j<i}\frac{\delta^{4}}{\delta\tilde{J}_{i}^{2}(z)\delta\tilde{J}_{j}^{2}(z)}\right)\right]\right\} ^{n}\Xi[\tilde{\mathbf{J}}]\nonumber \\
 & \approx\sum_{n=0}^{\infty}\frac{(-\tilde{\lambda})^{n}}{n!}\Biggl\{\prod_{k=1}^{n}\int d^{3}z_{k}\Bigl[M_{0}\bigl(\mathcal{F}[\tilde{J}_{3}]^{3}+3\mathcal{F}[\tilde{J}_{3}]\mathcal{F}'[\tilde{J}_{3}]\nonumber \\
 & \quad+\mathcal{F}[\tilde{J}_{3}](\mathcal{F}[\tilde{J}_{1}]^{2}+\mathcal{F}'[\tilde{J}_{1}])+\mathcal{F}[\tilde{J}_{3}](\mathcal{F}[\tilde{J}_{2}]^{2}+\mathcal{F}'[\tilde{J}_{2}])\bigr)\nonumber \\
 & \quad+\tfrac{1}{4}\Bigl(\sum_{i=1}^{3}(3(\mathcal{F}'_{i})^{2}+6\mathcal{F}'_{i}\mathcal{F}_{i}^{2}+\mathcal{F}_{i}^{4})\nonumber \\
 & \quad+2\sum_{j<i}(\mathcal{F}_{j}^{2}+\mathcal{F}'_{j})(\mathcal{F}_{i}^{2}+\mathcal{F}'_{i})\Bigr)\Bigr]\Biggr\}\Xi[\tilde{\mathbf{J}}],\label{eq:Z-exp-tilded}
\end{align}
where the auxiliary functions $\mathcal{F}_{i}[\tilde{J}_{i}(\mathbf{z})]$
and $\mathcal{F}'_{i}[\tilde{J}_{i}(\mathbf{z})]$ have already been
defined above.

This expression represents the exact combinatorial expansion in powers
of $\tilde{\lambda}$. However, when the propagator $Q_{i}(\mathbf{r},\mathbf{r}')$
is taken in the strictly local limit ($\tilde{\xi}\to0$), the functional
derivatives acting on $\mathcal{F}_{i}$ generate both local thermodynamic
terms and non-local connected terms (originating from the global spin
constraint). As demonstrated in Appendix A, these connected terms
scale as $\mathcal{O}(1/N_{{\rm eff}})$ relative to the dominant
disconnected terms, where $N_{{\rm eff}}$ is the macroscopic number
of spins within the polaron volume.

In the general non-local case (finite spin correlation length $\xi>0$)
the series remains {\em exact only up to first order} in $\lambda$
($n=1$). At second and higher orders, additional mixed Leibniz-rule
contributions appear when the operator $\mathcal{O}$ acts on the
$\mathcal{F}_{i}$ and $\mathcal{F}'_{i}$ generated by preceding
factors. These mixed terms are not captured by the simple product
form above and would require a considerably more elaborate treatment.

Two complementary and fully consistent extensions of the original
Dietl--Spałek Gaussian theory are therefore possible, each discussed
in detail in the subsequent sections: 
\begin{enumerate}
\item \textbf{Non-local BMP with first-order anharmonic correction} (valid
for any finite spin correlation length $\tilde{\xi}$). We retain
the exact non-local propagator $Q_{i}(\mathbf{r},\mathbf{r}')$ but
keep only the $n=1$ term in the expansion~(\ref{eq:Z-exp-tilded}). 
\item \textbf{Non-perturbative resummation of local BMP} ($\tilde{\xi}\to0$).
In the strict local limit relevant to the original DS theory, we can
construct a polaronic analogue of the Ginzburg criterion (Appendix
A). For a macroscopic polaron ($N_{{\rm eff}}\gg1$), the constraint-induced
non-localities are strongly suppressed. Neglecting contributions from
these minor contributions $\mathcal{O}(1/N_{{\rm eff}})$ allows us
to accurately reconstruct the infinite series of dominant local thermodynamic
fluctuations in exponential form. 
\end{enumerate}

\subsection{Non-local first-order correction}

We first consider the non-local case (path 1). Truncating the series~(\ref{eq:Z-exp-tilded})
at first order in $\tilde{\lambda}$ and setting $\tilde{\mathbf{J}}=\mathbf{0}$
(as required for $P(\boldsymbol{\Delta})$) yields the explicit correction
\begin{multline}
Z[\tilde{\mathbf{J}}=0]=\Xi[\tilde{\mathbf{J}}=0]\Biggl\{1-\tilde{\lambda}\int d^{3}z\Bigl[M_{0}\bigl(\mathcal{F}_{\parallel}^{3}+3\mathcal{F}_{\parallel}\mathcal{F}'_{\parallel}+\mathcal{F}_{\parallel}(\mathcal{F}_{\perp}^{2}+2\mathcal{F}'_{\perp})\bigr)\\
+\tfrac{1}{4}\Bigl(\sum_{i=1}^{3}(3(\mathcal{F}'_{i})^{2}+6\mathcal{F}'_{i}\mathcal{F}_{i}^{2}+\mathcal{F}_{i}^{4})+2\sum_{j<i}(\mathcal{F}_{j}^{2}+\mathcal{F}'_{j})(\mathcal{F}_{i}^{2}+\mathcal{F}'_{i})\Bigr)\Bigr]\Biggr\},\label{eq:Z-first-order-nonlocal-tilded}
\end{multline}
where $\mathcal{F}_{i}\equiv\mathcal{F}_{i}[\tilde{\mathbf{J}}=0]$
and $\mathcal{F}'_{i}\equiv\mathcal{F}'_{i}[\tilde{\mathbf{J}}=0]$
are the explicit building blocks given in Eq.~(\ref{eq:F-Fprime-zero-tilded}).

The probability distribution $P(\boldsymbol{\Delta})$ of the spin
splitting $\boldsymbol{\Delta}$ calculated to first order in the
anharmonicity parameter $\tilde{\lambda}$ for the non-local case
(finite spin correlation length $\tilde{\xi}$) reads 
\begin{equation}
\begin{aligned}P(\boldsymbol{\Delta}) & =2C\exp\!\Bigl(-\frac{\mathcal{H}_{1}}{k_{B}T}\Bigr)\cosh\!\left(s\frac{\Delta}{k_{B}T}\right)\Xi[\tilde{\mathbf{J}}=0]\\
 & \quad\times\Biggl\{1-\tilde{\lambda}M_{0}\int d^{3}z\bigl[\mathcal{F}_{\parallel}^{3}+3\mathcal{F}_{\parallel}\mathcal{F}'_{\parallel}+\mathcal{F}_{\parallel}(\mathcal{F}_{\perp}^{2}+2\mathcal{F}'_{\perp})\bigr]\\
 & \quad-\frac{\tilde{\lambda}}{4}\int d^{3}z\Bigl[\bigl(2\mathcal{F}'_{\perp}+\mathcal{F}'_{\parallel}+\mathcal{F}_{\perp}^{2}+\mathcal{F}_{\parallel}^{2}\bigr)^{2}\\
 & \quad+2\bigl(2(\mathcal{F}'_{\perp})^{2}+(\mathcal{F}'_{\parallel})^{2}\bigr)+4\bigl(\mathcal{F}'_{\perp}\mathcal{F}_{\perp}^{2}+\mathcal{F}'_{\parallel}\mathcal{F}_{\parallel}^{2}\bigr)\Bigr]\Biggr\}.
\end{aligned}
\label{eq:P-first-order-final-tilded}
\end{equation}
Here and in all subsequent expressions we adopt the following indexing
convention for the longitudinal (parallel) and transverse (perpendicular)
components with respect to the mean-field magnetization $\mathbf{M}_{0}$:
$\mathcal{F}_{\parallel}\equiv\mathcal{F}_{3}$, $\mathcal{F}'_{\parallel}\equiv\mathcal{F}'_{3}$
(parallel/longitudinal direction), while for the transverse plane
isotropy implies $\mathcal{F}_{\perp}^{2}\equiv F_{1}^{2}+F_{2}^{2}$
and $\mathcal{F}'_{1}=\mathcal{F}'_{2}\equiv\mathcal{F}'_{\perp}$.

Performing the spatial integrals over the non-local propagator $Q_{i}(\mathbf{r},\mathbf{r}')$
in the strict local limit $\tilde{\xi}\to0$ requires careful treatment
of the auxiliary functions $\mathcal{F}_{i}$ and $\mathcal{F}'_{i}$.
In this limit, the propagator approaches $Q_{i}(\mathbf{r},\mathbf{r}')\to\tilde{\mu}_{i}^{-1}\delta^{3}(\mathbf{r}-\mathbf{r}')$,
which formally cancels the mass dependence in the amplitude field
$\mathcal{F}_{i}(\mathbf{z})\propto\Lambda_{i}|\varphi(\mathbf{z})|^{2}$
while consistently regenerating the proper variance $\Phi_{i}$ in
the second derivative $\mathcal{F}'_{i}(\mathbf{z})\propto-\Phi_{i}|\varphi(\mathbf{z})|^{4}$.

By rigorously substituting these local operator limits back into Eq.~(\ref{eq:P-first-order-final-tilded})
and avoiding any premature isotropic approximations, we strictly preserve
the symmetry-breaking induced by the mean-field magnetization $M_{0}$.
To express the final probability distribution in terms of physical
observables, we now restore the explicit temperature dependence $\beta=1/k_{B}T$
and the original dimensional parameters of the Ginzburg-{}-Landau-{}-Wilson
functional. This yields the exact, fully anisotropic first-order local
distribution:

\begin{equation}
\begin{aligned}P(\boldsymbol{\Delta}) & =2C\exp\!\Bigl(-\frac{\mathcal{H}_{1}}{k_{B}T}\Bigr)\cosh\!\left(s\frac{\Delta}{k_{B}T}\right)\Xi[\mathbf{J}=0]\\
 & \quad\times\Biggl\{1-\frac{\lambda}{k_{B}T}\Biggl[\frac{1}{2}M_{0}\mathcal{I}_{3}\Lambda_{\parallel}\left(2\Lambda^{2}-3\Phi_{\parallel}-2\Phi_{\perp}\right)\\
 & \quad+\frac{1}{4}\mathcal{I}_{4}\left(\Lambda^{4}-\Lambda_{\parallel}^{2}(3\Phi_{\parallel}+2\Phi_{\perp})-\Lambda_{\perp}^{2}(\Phi_{\parallel}+4\Phi_{\perp})+\frac{3}{4}\Phi_{\parallel}^{2}+\Phi_{\parallel}\Phi_{\perp}+2\Phi_{\perp}^{2}\right)\Biggr]\Biggr\},
\end{aligned}
\label{eq:28-local}
\end{equation}
where we have introduced the effective vertex integrals $\mathcal{I}_{3}$
and $\mathcal{I}_{4}$. These incorporate the bare spatial overlap
integrals $\overline{I_{n}}\equiv\int d^{3}r\,|\varphi(\mathbf{r})|^{2n}$
scaled by the appropriate powers of the local propagator constraint
constant $c_{0}\equiv\frac{1}{2}\Bigl(\frac{\alpha}{g\mu_{B}}\Bigr)^{2}\overline{I_{2}}$:

\begin{equation}
\mathcal{I}_{n}\equiv\frac{\overline{I_{n}}}{c_{0}^{n}}\Bigl(\frac{\alpha}{g\mu_{B}}\Bigr)^{n}.
\end{equation}

This absorption is mathematically required because the functional
derivatives generating the local amplitudes $\mathcal{F}_{i}$ and
variances $\mathcal{F}'_{i}$ extract different powers of the un-dressed
constraint kernel. Consequently, the cubic fluctuations scale with
$c_{0}^{-3}$ and the quartic fluctuations with $c_{0}^{-4}$. This compact form is the strict local limit of the first-order non-local correction. In contrast to purely isotropic approximations, it explicitly
demonstrates how the host magnetization differentiates the thermodynamic
fluctuation variances into distinct longitudinal ($\Phi_{\parallel}$)
and transverse ($\Phi_{\perp}$) components.

The general non-local expression Eq.~(\ref{eq:P-first-order-final-tilded}
) retains the full spatial dependence of the spin correlations through
the Yukawa-like propagator $Q_{i}(\mathbf{r},\mathbf{r}')$ while
incorporating the leading cubic and quartic anharmonicities. It provides
a controlled perturbative extension of the Gaussian theory valid for
any finite $\xi$; explicit non-local convolutions for the hydrogenic
1s envelope function are derived in Appendix D. In the limit $\lambda\to0$
(for any finite $\xi$) it reduces to the non-local Gaussian theory.
The original Dietl--Spałek result is recovered exactly only upon
taking the additional strict local limit $\xi\to0$.

The explicit inclusion of the spatial spin correlation ($\xi>0$)
in the first-order perturbative theory yields profound physical consequences
for the polaron stability. The finite correlation length introduces
a spatial "rigidity" to the host magnetization, which effectively
resists the abrupt spatial collapse of the electron wavefunction.
For moderate exchange couplings (e.g., $J_{c}\approx150$ meV), this
non-local perturbative model performs remarkably well, predicting
a smooth self-trapping transition with the Bohr radius shrinking to
physical dimensions and the most probable spin splitting $\overline{\Delta}$
remaining well within the realistic sub-eV bounds down to temperatures
closely approaching $T_{c}$. However, numerical exploration of this
non-local first-order correction exposes its fundamental limit of
applicability. In the regime of strong exchange coupling ($J_{c}\gtrsim250$
meV) or in the immediate vicinity of the critical point $T\to T_{c}^{+}$,
the drastic volume contraction of the polaron overwhelmingly amplifies
the local variance of thermodynamic fluctuations. Consequently, the
effective perturbation parameter $\lambda_{p}$ ceases to be small.
The truncated polynomial expansion derived from the Ginzburg–Landau–Wilson
functional, $1-\lambda_{p}(u^{2}-5u+3.75)$, undergoes a mathematical
breakdown. Without a strictly bounding non-linear barrier, the thermodynamic
variational procedure pushes the most probable spin splitting to unphysical,
divergent values (greatly exceeding the semiconductor bandgap). This
dielectric-like catastrophe unambiguously demonstrates that the perturbative
treatment of non-Gaussian fluctuations is insufficient to capture
the deep self-trapping regime. To restore thermodynamic stability
and prevent the unphysical runaway of $\overline{\Delta}$, one must
step beyond the first-order expansion and perform a full non-perturbative
resummation of the anharmonic terms to recover the confining potential
barrier ($\propto-\Delta^{4}$). Nevertheless, as discussed in Appendix
A, an exact infinite resummation of spatially connected graphs for
a finite $\xi$ is analytically intractable. To resolve this, in the
following section (Sec. II.E), we take advantage of the macroscopic
nature of the polaron ($N_{\text{eff}}\gg1$) and proceed to the strict
local limit ($\xi\to0$). In this limit, the volume suppression of
connected graphs justifies the exact infinite resummation of the local
thermodynamic diagrams, ultimately rescuing the stability of the bound
magnetic polaron at arbitrary coupling strengths.

\subsection{Fully non-perturbative local resummation}

In the local limit (path 2) the full non-perturbative resummation
becomes possible (see Appendix A for more details). With $Q_{i}(\mathbf{r},\mathbf{r}')=\frac{1}{\tilde{\mu}_{i}}\delta^{3}(\mathbf{r}-\mathbf{r}')$
(and the standard ultraviolet regularization $q_{i}=0$) all operators
at different points can be considered commutative, and the infinite
series then sums exactly to 
\begin{equation}
Z[\tilde{\mathbf{J}}=0]=\Xi[\tilde{\mathbf{J}}=0]\,\exp\!\Bigl(-\tilde{\lambda}\int d^{3}z\,\mathcal{O}(\mathcal{F}_{i}(\mathbf{z}),\mathcal{F}'_{i}(\mathbf{z});M_{0})\Bigr),\label{eq:Z-resum-local-tilded}
\end{equation}
where the local operator $\mathcal{O}$ now contains \emph{both} the
cubic contribution (proportional to $M_{0}$) and the pure quartic
contribution: 
\begin{align}
\mathcal{O} & =M_{0}\Bigl[\mathcal{F}_{\parallel}^{3}+3\mathcal{F}_{\parallel}\mathcal{F}'_{\parallel}+\mathcal{F}_{\parallel}(\mathcal{F}_{\perp}^{2}+2\mathcal{F}'_{\perp})\Bigr]\nonumber \\
 & \quad+\frac{1}{4}\Bigl[\sum_{i=1}^{3}\bigl(3(\mathcal{F}'_{i})^{2}+6\mathcal{F}'_{i}\mathcal{F}_{i}^{2}+\mathcal{F}_{i}^{4}\bigr)\nonumber \\
 & \quad+2\sum_{j<i}(\mathcal{F}_{j}^{2}+\mathcal{F}'_{j})(\mathcal{F}_{i}^{2}+\mathcal{F}'_{i})\Bigr].\label{eq:O-operator-local-tilded}
\end{align}
Equation~(\ref{eq:Z-resum-local-tilded}) together with the local
operator~(\ref{eq:O-operator-local-tilded}) constitutes the desired
closed-form inclusion of the full GLW anharmonicity (expressed in
tilded parameters). Returning now to the original unrescaled physical
parameters to derive the probability distribution $P(\boldsymbol{\Delta})$
and the total free energy $\Delta F$, we define the effective dimensionless
anharmonicity parameter as 
\begin{equation}
\lambda_{p}\equiv\lambda\,\mathcal{I}_{4}\,\varepsilon_{p}^{2},
\end{equation}
where $\lambda$ is the coefficient of the quaternary term of the
GLW functional (eq.~(\ref{eq:GLW_functional})), and all subsequent
expressions contain the proper combination $\lambda_{p}k_{B}T$ (or
$\lambda_{p}\varepsilon_{p}$) that guarantees dimensionless exponents.

By performing a change of variables to the modulus $\Delta$ and the
angle $\theta$ between $\boldsymbol{\Delta}$ and the uniform field
$\boldsymbol{\Delta}_{0}$ (where $x=\cos\theta$), we arrive at the
central result of our formal treatment. This stabilized distribution,
which bridges the microscopic carrier-spin exchange with the fully
resummed non-Gaussian fluctuations of the host, is obtained by synthesizing
the general form Eq.~(\ref{eq:P-Delta-2}) with the resummation expression~(\ref{eq:Z-resum-local-tilded})
and the anharmonic operator~(\ref{eq:O-operator-local-tilded}):
\begin{equation}
\begin{aligned}P(\Delta) & =\mathcal{N}\exp\left(-\frac{\mathcal{H}_{1}}{k_{B}T}\right)\cosh\left(\frac{s\Delta}{k_{B}T}\right)\left[\Xi[0]\exp\left(-\frac{\lambda}{k_{B}T}\int d^{3}z\,\mathcal{O}(z;M_{0})\right)\right]\\
 & =N\Delta^{2}\cosh\left(\frac{s\Delta}{k_{B}T}\right)I_{\text{ang}}(\Delta;\Delta_{0},\lambda_{p}),
\end{aligned}
\label{eq:final_P_delta}
\end{equation}
where $\mathcal{N}$ is the normalization constant satisfying $\int_{0}^{\infty}P(\Delta)\,d\Delta=1$
and $I_{{\rm ang}}$ is the angular integral over the direction of
$\Delta$, given in closed form by the error function $\operatorname{erf}$
or the imaginary error function $\operatorname{erfi}$ (algebraic
details of angular integration, the resulting expressions, and the
special limiting cases are presented in Appendix~B).

For notational convenience we introduce the auxiliary dimensionless
variable 
\begin{equation}
u=\frac{\Delta^{2}}{8k_{B}T\varepsilon_{p}}
\end{equation}
and its static counterpart $u_{0}=\Delta_{0}^{2}/(8k_{B}T\varepsilon_{p})$.
In this local limit the Gaussian part of the generating functional
involves the quantity $\Phi=2k_{B}T\varepsilon_{p}$, where the characteristic
polaron energy is 
\begin{equation}
\varepsilon_{p}=\frac{\alpha^{2}\chi}{32\pi a_{B}^{3}(g\mu_{{\rm B}})^{2}}.
\end{equation}

\subsection{Symmetric Paramagnetic Case}

The most significant achievement of this non-perturbative theoretical
extension is the rigorous description of the bound magnetic polaron
in the symmetric paramagnetic phase ($T>T_{c}$). In this regime,
in the absence of an external magnetic field ($\mathbf{B}_{ext}=0$),
the macroscopic magnetization of the host vanishes ($\mathbf{M}_{0}=0$).
This directly dictates that the uniform macroscopic spin splitting
is also zero ($\mathbf{\Delta}_{0}=0$).

In this symmetric limit, the mathematical framework simplifies considerably.
Specifically, the angular integration coefficients from Appendix B
reduce exactly to $A=B=0$. Consequently, the angular integral in
Eq. (\ref{eq:final_P_delta}) reduces to a compact closed-form expression,
yielding a stabilized probability distribution for the internal spin
splitting:

\begin{equation}
P(\Delta)=\mathcal{N}\,4\pi\Delta^{2}\cosh\left(\frac{s\Delta}{k_{B}T}\right)\exp\left\{ -u-\lambda_{p}k_{B}T\Bigl(u^{2}-5u+\tfrac{15}{4}\Bigr)\right\} ,\label{eq:Pparamag}
\end{equation}
where $\mathcal{N}$ is again chosen so that $\int_{0}^{\infty}P(\Delta)\,d\Delta=1$.
This expression directly generalizes the Gaussian Dietl--Spałek result
and remains stable even when the magnetic susceptibility diverges
at $T_{c}$.

Based on this probability distribution, we can calculate the most
probable value of the spin splitting $\overline{\Delta}$ (the value
that maximizes $P({\Delta})$). It is obtained from the stationarity
condition $\frac{d}{d\Delta}\ln P(\Delta)=0$. The logarithm (up to
an irrelevant additive constant) reads: 
\begin{equation}
\ln P(\Delta)=2\ln\Delta+\ln\cosh\left(\frac{s\Delta}{k_{B}T}\right)-u-\lambda_{p}k_{B}T\left(u^{2}-5u+\frac{15}{4}\right).
\end{equation}
Differentiating with respect to $\Delta$ and setting the derivative
to zero yields the quartic-transcendental equation: 
\begin{equation}
\frac{\lambda_{p}}{4\varepsilon_{p}}\overline{\Delta}^{4}+\left[1-5\lambda_{p}k_{B}T\right]\overline{\Delta}^{2}-4s\varepsilon_{p}\overline{\Delta}\tanh\left(\frac{s\overline{\Delta}}{k_{B}T}\right)-8\varepsilon_{p}k_{B}T=0.
\end{equation}

\textbf{Gaussian limit} ($\lambda_{p}=0$): If the anharmonicity parameter
is switched off, the above stationarity condition immediately simplifies
to 
\begin{equation}
\overline{\Delta}^{2}-4s\varepsilon_{p}\overline{\Delta}\tanh\left(\frac{s\overline{\Delta}}{k_{B}T}\right)-8\varepsilon_{p}k_{B}T=0,
\end{equation}
recovering exactly the most-probable spin splitting of the original
Dietl--Spałek theory (for $s=1/2$).

This equation can be solved numerically for $\overline{\Delta}(T)$
(or perturbatively for small $\lambda_{p}$) to quantify how the quartic
stabilizing term modifies the typical spin splitting inside the BMP
even in the paramagnetic phase.\textbf{ }

\textbf{Free energy and stationary states}

For the fully resummed case, the total free energy of the BMP is constructed
using the angularly integrated probability distribution $P(\boldsymbol{\Delta})$
(Appendix~B). The renormalized Schrödinger equation for the stationary
envelope $\varphi(\mathbf{r})$ follows from the Euler--Lagrange
variation of this total free-energy functional 
\begin{equation}
\Delta F[\varphi]=\langle\varphi|\mathcal{H}_{1}|\varphi\rangle-k_{B}T\ln\left(\int_{0}^{\infty}\Delta^{2}\,d\Delta\,\cosh\left(s\beta\Delta\right)I_{{\rm ang}}(\Delta;\Phi[\varphi])\right)
\end{equation}
under the normalization constraint $\int|\varphi(\mathbf{r})|^{2}\,d^{3}r=1$,
where $I_{{\rm ang}}(\Delta;\Phi)$ is the angular integral given
in the appendix (involving $\operatorname{erf}$ or $\operatorname{erfi}$).
Because $I_{{\rm ang}}$ depends on $\Phi[\varphi]$ through the coefficients
$A(\Phi)$, $B(\Phi)$, and $C(\Phi)$, the functional derivative
$\delta\Delta F/\delta\varphi^{*}(\mathbf{r})$ generates additional
non-linear terms. The resulting stationarity condition reads formally
\begin{equation}
\mathcal{H}_{1}\varphi-s^{2}\beta\frac{\delta\Phi[\varphi]}{\delta\varphi^{*}}-k_{B}T\frac{\delta}{\delta\varphi^{*}}\ln\left(\int_{0}^{\infty}\Delta^{2}\cosh(s\beta\Delta)I_{{\rm ang}}(\Delta;\Phi)\,d\Delta\right)-E\varphi=0.\label{eq:Sch-resummed}
\end{equation}
The second functional derivative (with respect to $\Phi$) acts on
the full resummed partition function and cannot be reduced to a simple
algebraic operator as in the Gaussian or first-order cases; it must
in general be evaluated numerically. In the perturbative local limit
(small $\lambda$) Eq.~(\ref{eq:Sch-resummed}) reduces to the explicit
analytical form, 
\begin{equation}
\mathcal{H}_{1}\varphi-s^{2}\beta\frac{\delta\Phi[\varphi]}{\delta\varphi^{*}}-k_{B}T\frac{\delta}{\delta\varphi^{*}}\ln I[\varphi^{*}]-E\varphi=0,
\end{equation}
with the correction factor 
\begin{equation}
I[\varphi^{*}]\equiv2+4(s\beta)^{2}\Phi[\varphi^{*}]-2\lambda_{p}k_{B}T\,(s\beta)^{4}\Phi[\varphi^{*}]^{2}\bigl\{2(s\beta)^{2}\Phi[\varphi^{*}]+5\bigr\}.
\end{equation}
Thus the non-perturbative local resummation provides the exact thermodynamic
stabilization while the non-local first-order treatment offers a practical
route for finite-correlation-length systems.

Once the optimal orbital size $a_{B}$ is established through this
variational procedure, the most-probable internal spin splitting $\overline{\Delta}$
is extracted directly by maximizing the probability distribution $\ln P(\Delta)$
for the optimized envelope.

While $\Delta F[\varphi]$ represents the total integrated free energy
of the polaron state necessary for the variational optimization of
the orbital, understanding the local symmetry breaking requires analyzing
the effective free-energy landscape for a given fluctuation magnitude
$\Delta$, defined via the probability distribution as $F_{\text{eff}}(\Delta)=-k_{B}T\ln P(\Delta)$
and the proper Landau thermodynamic potential $V_{\text{eff}}(\Delta)$.
The transformation of the probability density from the three-dimensional
vector space of magnetization fluctuations to the scalar magnitude
$\Delta$ introduces a phase-space volume factor (Jacobian) proportional
to $\Delta^{2}$. In the energy representation, this generates a purely
statistical entropy term $-2k_{B}T\ln\Delta$, which forces $F_{\text{eff}}(\Delta)\to\infty$
as $\Delta\to0$. While this geometrical phase-space term is strictly
required to determine the statistically most probable spin splitting
$\overline{\Delta}$ by maximizing $P(\Delta)$, it does not represent
a physical repulsive force acting on the electron. The actual thermodynamic
landscape experienced by the carrier is governed by the reduced Landau
potential:

\begin{equation}
V_{\text{eff}}(\Delta)=F_{\text{eff}}(\Delta)+2k_{B}T\ln\Delta.\label{eq:L_potential}
\end{equation}

As the temperature decreases towards the critical regime, the spin-exchange
interaction continuously destabilizes the trivial parabolic minimum
at $\Delta=0$, morphing the energetic landscape into a characteristic
``Mexican hat'' profile. The local symmetry is spontaneously broken,
and the electron is driven into a new, stable potential well at finite
$\Delta$, which is strictly prevented from collapsing by the quartic
stabilization term $\propto\lambda_{p}\Delta^{4}$.

It is instructive to place the mechanism of deep magnetic self-trapping presented here within the broader context of the classical bound magnetic polaron theory formulated by Dietl and Spałek. In their pioneering work, DS elegantly analyzed the weak-coupling regime typical for diluted magnetic semiconductors (such as CdMnSe). They explicitly restricted their model to the "large polaron" limit, where the magnetic exchange energy is strictly much smaller than the electrostatic binding energy ($\epsilon_p \ll E_D$). Under these well-defined conditions, the donor orbital radius remains essentially rigid (yielding a negligible $\sim 2\%$ correction), and the authors rigorously demonstrated that the phase transition predicted by the mean-field approximation is completely washed out by Gaussian thermodynamic fluctuations.  Our non-perturbative treatment serves as a natural generalization of the DS framework, extending its applicability into the strong-coupling and deep self-trapping regimes. When the carrier-spin exchange interaction is strong—such as in ferromagnetic semiconductors or highly coupled paramagnetic hosts—the magnetic energy gain can no longer be treated as a small perturbation. To maximize the local exchange interaction, the electron undergoes a spontaneous and pronounced spatial contraction. Because the exchange energy gain scales as $1/a_B^3$ while the kinetic energy penalty scales only as $1/a_B^2$, pushing the purely Gaussian model into this unconstrained shrinking regime would eventually trigger an unphysical spatial collapse.

Let us examine the stability loss condition at $\Delta=0$ by studying
the convexity of the effective Landau potential $V_{\text{eff}}(\Delta)$.
In the isotropic paramagnetic phase ($\Delta_{0}=0$), substituting
the probability distribution given by Eq.~(\ref{eq:final_P_delta})
and omitting constant terms, we obtain:

\begin{equation}
V_{\text{eff}}(\Delta)=-k_{B}T\ln\cosh\left(\frac{s\Delta}{k_{B}T}\right)+\frac{\Delta^{2}}{8\varepsilon_{p}}+\lambda_{p}(k_{B}T)^{2}\left(\frac{\Delta^{4}}{64(k_{B}T\varepsilon_{p})^{2}}-\frac{5\Delta^{2}}{8k_{B}T\varepsilon_{p}}+\frac{15}{4}\right).
\end{equation}

To analytically determine the destabilization temperature $T_{\text{char}}$
where this local symmetry breaking occurs, we evaluate the convexity
of the effective Landau potential. Expanding $V_{\text{eff}}(\Delta)$
in a Taylor series around $\Delta=0$, using the approximation $\ln\cosh(x)\approx x^{2}/2$,
yields:

\begin{equation}
V_{\text{eff}}(\Delta)\approx\left[-\frac{s^{2}}{2k_{B}T}+\frac{1}{8\varepsilon_{p}}-\frac{5\lambda_{p}k_{B}T}{8\varepsilon_{p}}\right]\Delta^{2}+\mathcal{O}(\Delta^{4}).
\end{equation}

The trivial minimum at $\Delta=0$ becomes destabilized when the coefficient
of $\Delta^{2}$ becomes zero. Equating it to zero and multiplying
by $8\varepsilon_{p}k_{B}T$ gives:

\begin{equation}
-4s^{2}\varepsilon_{p}+k_{B}T(1-5\lambda_{p}k_{B}T)=0.
\end{equation}

For a donor electron, the spin is $s=1/2$, so $4s^{2}=1$. The equation
for the destabilization temperature $T_{\text{char}}$ then takes
the form:

\begin{equation}
-\varepsilon_{p}+k_{B}T-5\lambda_{p}(k_{B}T)^{2}=0.
\end{equation}

In the purely Gaussian limit ($\lambda_{p}=0$) inherent to the original DS theory, this recovers exactly the phase transition temperature $T=\varepsilon_{p}/k_{B}\equiv T_{p}$. In our extended quartic model ($\lambda_{p}>0$), solving the quadratic equation under the assumption that $\lambda_{p}$ is a small perturbative parameter gives:

\begin{equation}
T_{\text{char}}\approx T_{p}(1+5\lambda_{p}\varepsilon_{p}).
\end{equation}

This analytical result elegantly illuminates the fate of the mean-field instability. The apparent complete smearing of the phase transition, which is physically perfectly justified in the strict Gaussian large-polaron limit, naturally evolves when the strong-coupling regime is entered. The phase transition—manifesting itself physically as magnetic self-trapping—fundamentally exists. The quartic fluctuations not only shift the transition temperature upwards by $5\lambda_{p}\varepsilon_{p}^{2}/k_{B}$, but, when rigorously incorporated alongside the variational optimization of the electron's orbit, they provide the exact restoring force required to halt the aforementioned unphysical BMP collapse. The quartic term ($\lambda_p\Delta^4$) curls the effective Landau potential upwards, stabilizing a non-trivial local minimum.

Thus, our theory smoothly bridges the gap between the weakly coupled, large-polaron regime—excellently captured by the Gaussian DS model—and the strongly coupled regime. We demonstrate that what acts as a smeared crossover at weak coupling evolves into a true local phase transition, manifesting physically as the formation of a thermodynamically stable, self-trapped ferromagnetic droplet when the coupling threshold is crossed.

\section{Application of the model to ferromagnetic semiconductors}

As discussed in Section II.B, our previous renormalized-Gaussian approach
fixed the donor radius at $a_{B}$, precluding the study of magnetic
self-trapping. By employing the fully variationally optimized non-perturbative
theory developed above, we now apply this framework to the ferromagnetic
semiconductor GdN ($T_{c}\approx55$\,K). The microscopic parameters
entering the Ginzburg--Landau--Wilson functional are taken directly
from our earlier renormalized-Gaussian study \cite{Bednarski2014NJP}.
Specifically, the inverse susceptibility follows the Curie--Weiss
form $\mu=(T-T_{c})/C_{{\rm Curie}}$ with the experimental Curie
temperature $T_{c}=55\,\text{K}$, the quartic coefficient is $\lambda=1/n_{0}^{3}$
where $n_{0}=32$, and the calculations are performed in the strict
local limit $\xi\to0$. The unit cell of GdN (rocksalt structure)
contains 4 Gd$^{3+}$ ions with spin $S=7/2$, and the lattice constant
is $a=0.5\,\text{nm}$. The bare donor Bohr radius $a_{B0}=1.41\,\text{nm}$
and the effective Rydberg energy $Ry^{*}=127.5\,\text{meV}$ are likewise
taken from Ref.~\cite{Bednarski2014NJP}. The carrier--spin exchange
integral $J_{c}$ is treated as a tunable material parameter. In Ref.~\cite{Bednarski2014NJP}
the value $J_{c}=190\,\text{meV}$ provided an excellent description
of the experimental magnetization and carrier-concentration data for
GdN; \textit{ab initio} estimates reach up to $\sim350\,\text{meV}$
\cite{Sharma_2006}. Here we explore the full physically relevant
range $J_{c}=0$--$400\,\text{meV}$, thereby demonstrating the robustness
of the quartic stabilization and the sensitivity of the BMP properties
to the exchange strength.

The total free-energy functional 
\begin{equation}
\Delta F[\varphi,\Delta_{0}^{ext}]=\langle\varphi|\mathcal{H}_{1}|\varphi\rangle+F_{{\rm glw}}(M_{0})+\Delta F_{{\rm mag}}(\Delta_{0}^{ext};\Phi[\varphi])
\end{equation}
is minimized variationally with respect to the envelope radius $a_{B}$
(via the hydrogenic trial function) for a given uniform macroscopic
spin splitting $\Delta_{0}^{ext}$ (which vanishes in the paramagnetic
phase without an external magnetic field). The variational minimization
of the total free energy is performed using the analytical building
blocks for the 1s hydrogenic wavefunction derived in Appendix D, ensuring
both high numerical precision and computational efficiency. Here $F_{{\rm glw}}$
is the mean-field contribution per donor and $\Delta F_{{\rm mag}}$
is the fluctuation correction obtained from the fully resummed partition
function $Z[\Delta_{0}]$ (Eq.~(\ref{eq:Z-radial-general})).

To elucidate the fundamental mechanism of magnetic self-trapping and
the interplay between local fluctuations and macroscopic ordering,
we first analyze the variational properties of the bound magnetic
polaron as a function of the external macroscopic spin splitting $\Delta_{0}^{ext}$.
This macroscopic splitting is directly proportional to the static
uniform magnetization $M_{0}$ and can be driven by an external magnetic
field $B_{ext}$.

\begin{figure}[htbp]
\centering \includegraphics[width=0.75\textwidth]{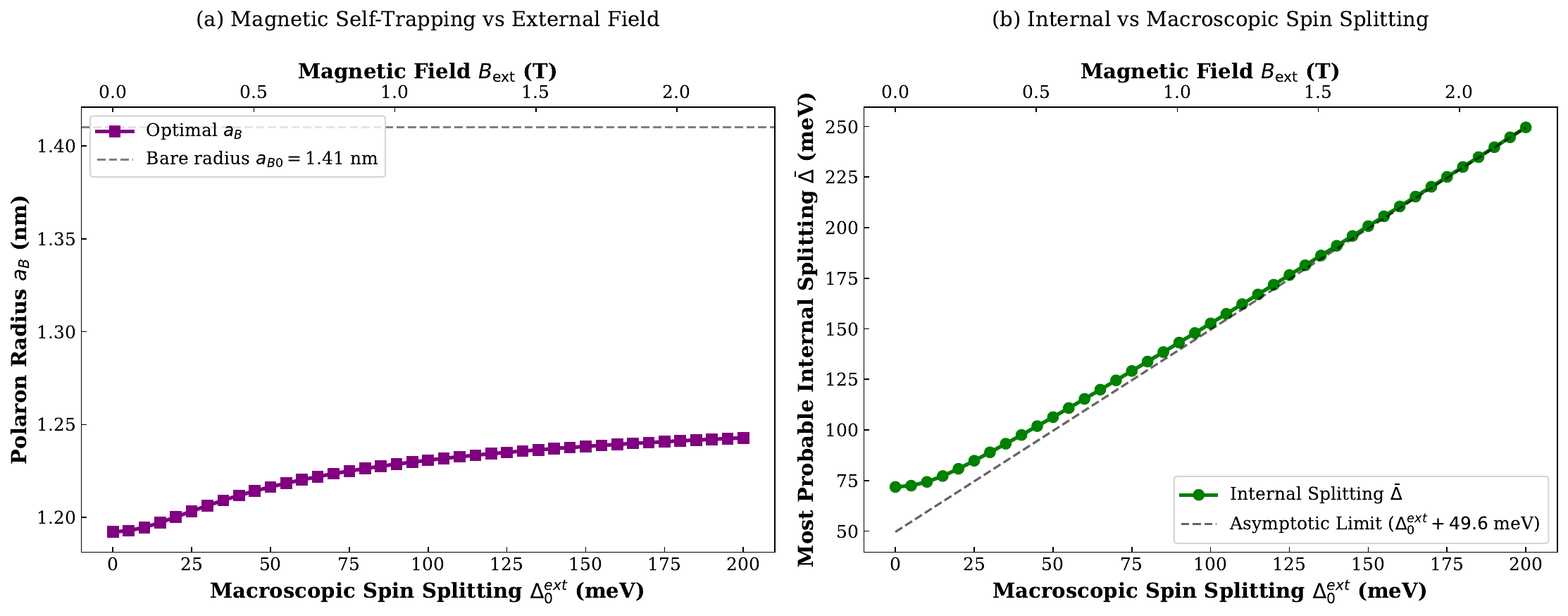}
\caption{Response of the bound magnetic polaron to the external macroscopic spin splitting $\Delta_{0}^{\text{ext}}$ at $T = 70$~K for the exchange integral $J_{c} = 190$~meV, with the equivalent external magnetic field $B_{\text{ext}}$ shown on the top axis. (a) Optimal polaron radius $a_{B}$ as a function of $\Delta_{0}^{\text{ext}}$, showcasing orbital relaxation and expansion toward the bare hydrogenic limit $a_{B0} = 1.41$~nm under strong fields. (b) Most probable internal spin splitting $\bar{\Delta}$ as a function of $\Delta_{0}^{\text{ext}}$. The dashed line indicates the strong-field asymptotic tracking limit given by $\bar{\Delta} = \Delta_{0}^{\text{ext}} + 49.6$~meV. The pronounced deviation at weak fields highlights the crucial role of thermodynamic fluctuations.}
\label{fig:BMP_Response} 
\end{figure}

Figure~\ref{fig:BMP_Response}(a) displays the optimal polaron radius
$a_{B}$ versus $\Delta_{0}^{ext}$. At zero macroscopic field, the
donor electron orbit is contracted ($a_{B}\approx1.19$\,nm) due
to magnetic self-trapping, as the electron must tightly localize to
effectively polarize the surrounding host spins. As the macroscopic
field increases, it enforces collinear alignment of the localized
spins throughout the crystal. Because the spins are already polarized
by the external field, the necessity for the electron to maintain
a highly localized state to gain exchange energy is lifted. Consequently,
the orbit relaxes and expands towards a saturated, weakly field-dependent
limit ($a_{B}\approx1.24$\,nm for the chosen parameters).

Figure~\ref{fig:BMP_Response}(b) illustrates the response of the
most probable internal spin splitting $\bar{\Delta}$ to the external
macroscopic field $\Delta_{0}^{ext}$. In the strong-field regime
($\Delta_{0}^{ext}\gg50$\,meV), the localized host spins are heavily
polarized along the external field direction, strongly suppressing
transverse fluctuations. In this limit, the internal spin splitting
simply tracks the macroscopic field with a constant longitudinal offset
($\bar{\Delta}\approx\Delta_{0}^{ext}+\Delta_{\text{offset}}$), where
$\Delta_{\text{offset}}\approx50$\,meV represents the saturated
local exchange field generated by the relaxed donor envelope ($a_{B}\approx1.24$\,nm).

However, a pronounced deviation from this asymptotic linear behavior
occurs in the weak-field regime ($\Delta_{0}^{ext}\to0$). Here, the
macroscopic field is insufficient to enforce collinearity, and the
physics is entirely dominated by the isotropic thermodynamic fluctuations
of the localized spins. Because the internal splitting $\bar{\Delta}$
reflects the magnitude of the local effective exchange field (which
is a vector sum of these unpolarized local fluctuations), it remains
large and finite even when $\Delta_{0}^{ext}=0$. The geometric ``bulge''
where the calculated $\bar{\Delta}$ exceeds the high-field asymptotic
limit is a direct, quantitative measure of the non-linear, fluctuation-driven
magnetic self-trapping inherent to the bound magnetic polaron.

\begin{figure}[htbp]
\centering \includegraphics[width=0.55\textwidth]{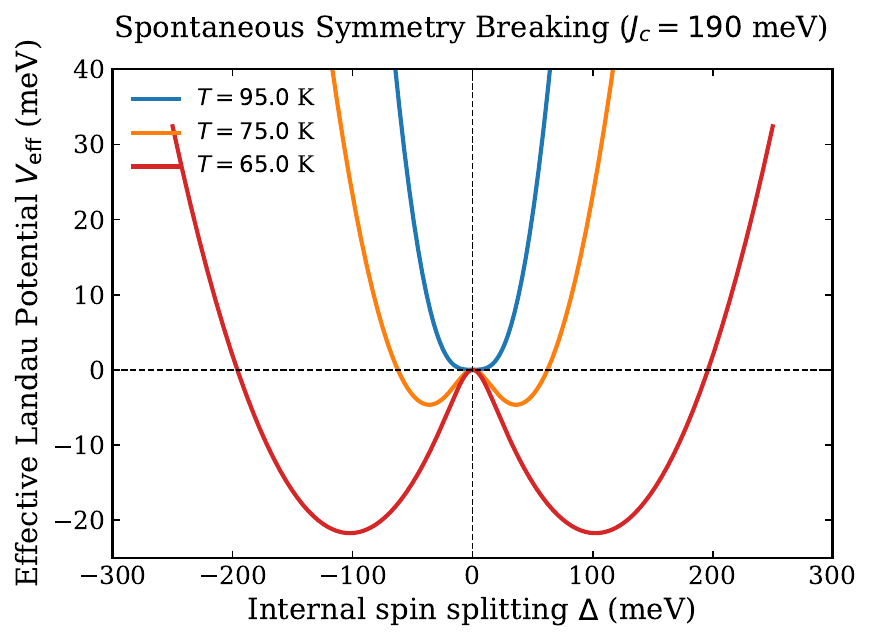}
\caption{The effective Landau thermodynamic potential $V_{\text{eff}}(\Delta)$
of the bound magnetic polaron at zero external field ($\Delta_{0}^{{\rm ext}}=0$)
and for $J_{{\rm c}}=190\,\text{meV}$ . As temperature decreases,
the trivial minimum at $\Delta=0$ destabilizes, and the quartic anharmonicities
stabilize a “Mexican hat” profile, vividly illustrating the spontaneous
local symmetry breaking and the emergence of a finite internal spin
splitting.}
\label{fig:Mexican_Hat} 
\end{figure}

A central prediction of the present theory is the spontaneous magnetic
self-trapping of the donor electron. In the paramagnetic phase ($T>T_{c}$)
and zero external magnetic field, the macroscopic splitting vanishes
($\Delta_{0}^{{\rm ext}}=0$) and coefficients A and B vanish. To
explicitly visualize the thermodynamic mechanism driving this zero-field
localization, we examine the proper Landau thermodynamic potential
defined in Eq. (\ref{eq:L_potential}), $V_{\text{eff}}(\Delta)=F_{\text{eff}}(\Delta)+2k_{B}T\ln\Delta,$where
$F_{\text{eff}}(\Delta)=-k_{B}T\ln P(\Delta)$ is the effective radial
free-energy profile. Subtracting the purely geometrical phase-space
entropy term ($-2k_{B}T\ln\Delta$, originating from the spherical
Jacobian) reveals the bare energetic landscape experienced by the
carrier. As illustrated in Fig.~\ref{fig:Mexican_Hat}, lowering
the temperature naturally destabilizes the trivial paramagnetic state,
allowing the quartic anharmonicities to sculpt the free energy into
a characteristic ``Mexican hat'' profile. 

\begin{figure}[htbp]
\centering \includegraphics[width=0.75\textwidth]{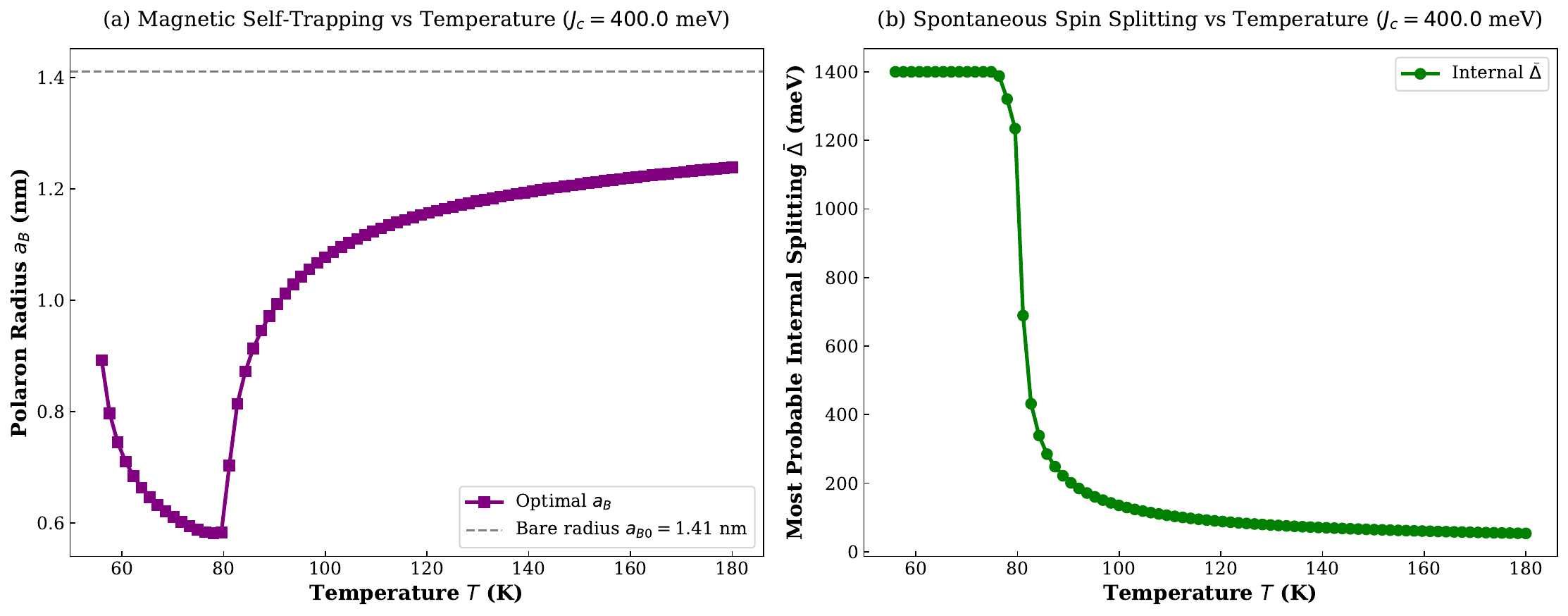}
\hfill{}\caption{Temperature dependence of the bound magnetic polaron properties in the paramagnetic phase ($\Delta_{0}^{\text{ext}} = 0$) calculated for a realistic exchange integral $J_{c} = 400.0$~meV. (a) Optimal polaron radius $a_{B}$ versus temperature, revealing a pronounced spatial contraction and a characteristic kink at $T_{\text{char}} \approx 79$~K due to the onset of non-Gaussian fluctuations. (b) Most probable internal spin splitting $\bar{\Delta}$ versus temperature, demonstrating the persistence of a robust local ferromagnetic order deep into the paramagnetic phase up to the effective ordering temperature $T^* \approx 155$--$160$~K.}
\label{fig:T_evolution} 
\end{figure}

By optimizing the total free-energy functional with respect to the
variational Bohr radius $a_{B}$, the model captures this delicate
competition between the kinetic-energy cost and the gain in fluctuation
energy upon orbital contraction. The strength of the resulting local
ordering is quantified by the most probable internal spin splitting
$\overline{\Delta}$. Figure~\ref{fig:T_evolution} shows the temperature dependence of the optimal polaron radius $a_{B}(T)$
for the realistic exchange integral $J_{c}=400\,\text{meV}$ (consistent
with \textit{ab initio} estimates). At high temperatures ($T\gtrsim160\,\text{K}$)
the polaron remains weakly bound with $a_{B}\approx a_{B0}=1.41\,\text{nm}$.
As temperature is lowered, $a_{B}$ decreases drastically, reaching
a pronounced minimum near $T\approx79\,\text{K}$. This kink marks
a crossover between two competing branches of the free-energy landscape.
Starting from the bulk Curie temperature $T_c \approx 55$ K and increasing temperature, a polaronic tail develops. 
In the regime between $T_c$ and $T_{char}$, the polaron undergoes severe spatial contraction (Fig. 3(a)) to maximize the exchange energy gain from local fluctuations. As a direct result of this deep self-trapping, the enclosed spins remain strongly polarized, maintaining a saturated and constant internal spin splitting $\overline{\Delta}$ (Fig. 3(b)) even though the macroscopic uniform field vanishes ($\Delta_0^{ext} = 0$).
At the characteristic temperature $T_{char} \approx 79$ K, the kinetic energy cost and the anharmonic (quartic) thermodynamic fluctuations provide a strong enough restoring force to counteract further orbital collapse. Because this crossover occurs in the symmetric paramagnetic phase ($\Delta_0^{ext} = 0$), the angular integration coefficients $A$ and $B$ (Appendix B) vanish identically. The observed kink in $a_B(T)$ is therefore a strictly radial phenomenon, marking the exact thermodynamic balance point within the one-dimensional "Mexican hat" free-energy landscape.
Above $T_{char}$, thermal fluctuations begin to overcome the self-trapping potential, causing the polaron orbit to gradually expand. This spatial swelling rapidly weakens the effective local exchange field (which scales as $1/a_B^3$), leading to a sharp drop in the most probable internal spin splitting $\overline{\Delta}$. Nevertheless quartic stabilization ensures that $\overline{\Delta}$ remains finite and sizable up to the effective polaron ordering temperature $T^* \approx 155-160$ K, demonstrating thermodynamically stable, ferromagnetically ordered BMPs well above the bulk $T_c$.

\begin{figure}[htbp]
\centering \includegraphics[width=0.75\textwidth]{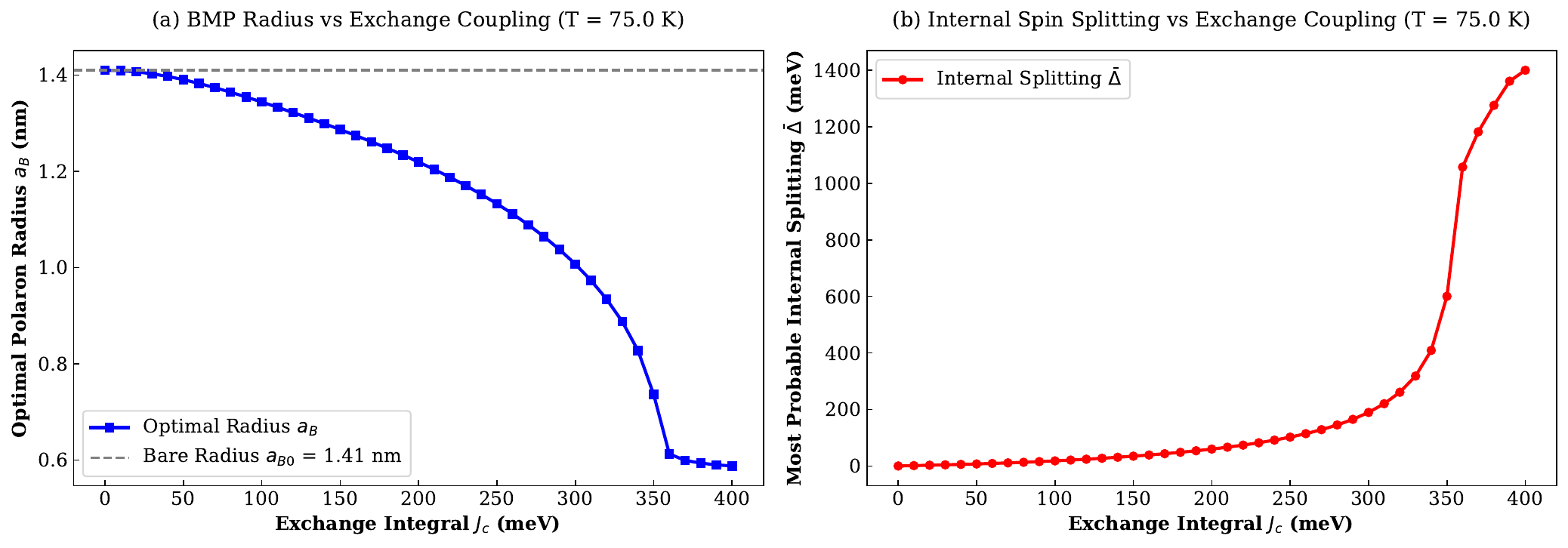}
\hfill{}\caption{Variational properties of the bound magnetic polaron as a function of the carrier-spin exchange integral $J_{c}$ evaluated at a fixed temperature $T = 75.0$~K in the paramagnetic phase. (a) Optimal polaron radius $a_{B}$ versus $J_{c}$, showcasing the transition from a large polaron regime to a sudden orbital collapse. (b) Most probable internal spin splitting $\bar{\Delta}$ versus $J_{c}$. The sharp threshold behavior around $J_{c} \approx 330$~meV signals the cooperative collapse into a highly localized, self-trapped small polaron state.}
\label{fig:Jc_dependence} 
\end{figure}

The profound role of the non-linear positive feedback between the
electron localization and local spin polarization is best illustrated
by the dependence on the exchange integral $J_{c}$, shown in Fig.~\ref{fig:Jc_dependence}
at fixed $T=75\,\text{K}$. For $J_{c}<400\,\text{meV}$, a large
polaron regime is observed, where a gradual decrease of $a_{B}$ is
accompanied by a slow, quasi-linear growth of the internal splitting
$\bar{\Delta}$. However, above a critical coupling strength ($J_{c}\approx330\,\text{meV}$),
the system undergoes a drastic polaron collapse. Because the local
exchange field scales as $1/a_{B}^{3}$, the orbital contraction rapidly
amplifies the local magnetization, which in turn deepens the trapping
potential. This leads to a highly localized state ($a_{B}\approx0.6\,\text{nm}$) with saturated internal spin splitting exceeding $1\,\text{eV}$. 

\begin{figure}[htbp]
\centering \includegraphics[width=0.8\textwidth]{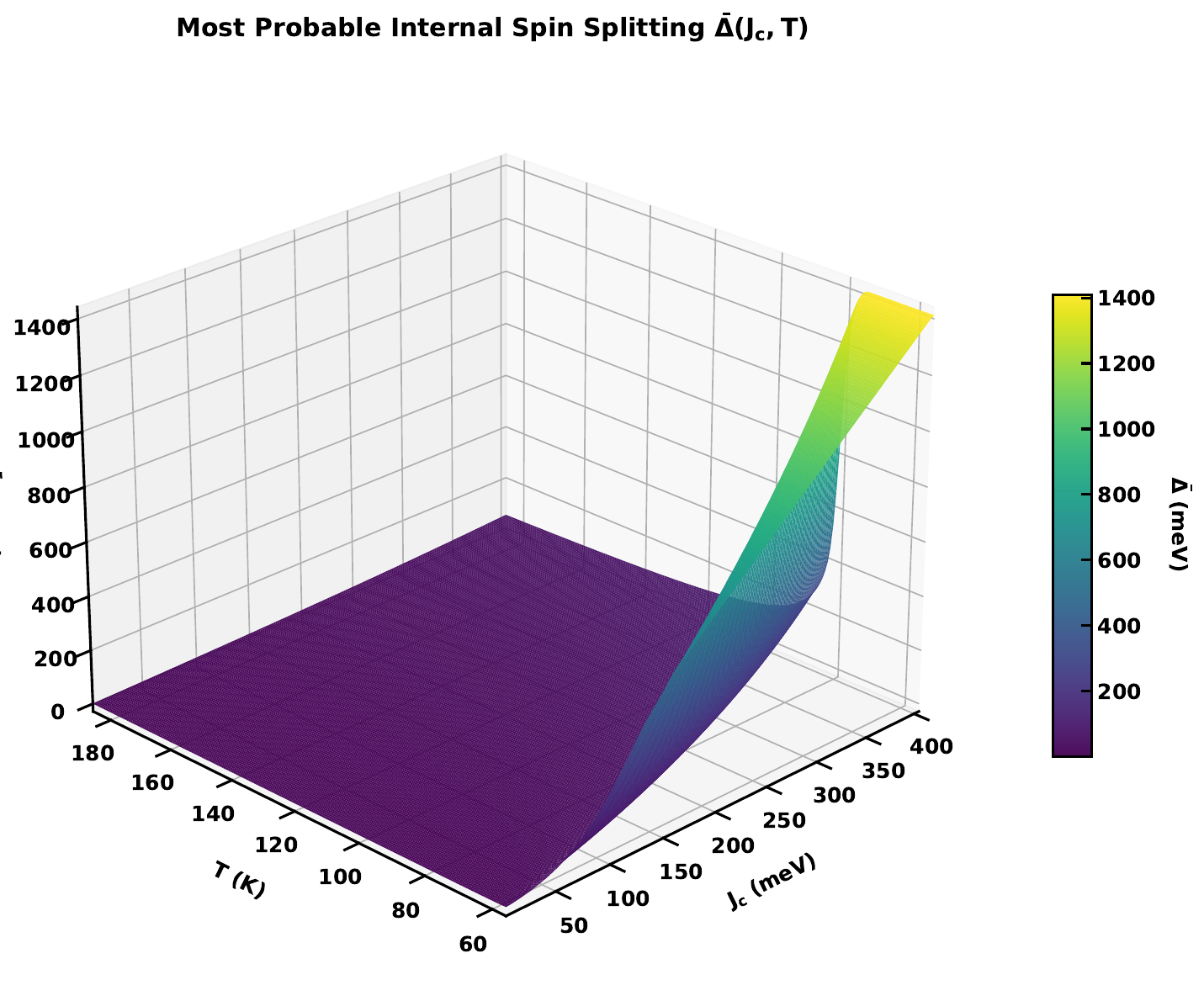}
\caption{Three-dimensional surface plot of the most probable internal spin
splitting $\bar{\Delta}(J_{c},T)$ obtained from the fully resummed
variationally optimized theory. The non-zero elevation demonstrates
the persistence of local ferromagnetic ordering well into the paramagnetic
phase despite the macroscopic magnetization being zero.}
\label{fig:surface_Delta0} 
\end{figure}

It is worth addressing here the physical consistency of the model
in the extreme strong-coupling regime ($J_{c}\gtrsim330$ meV), where
the optimal polaron radius is drastically reduced to $a_{B}\approx0.6$
nm. At first sight, such strong localization might appear to invalidate
the continuous-medium approximation. However, the actual microscopic
structure of defects in GdN must be taken into account. The shallow
donors responsible for polaron formation in this material are nitrogen
vacancies ($V_{N}$). Placing the center of the hydrogen-like wave
function at an anion site in the rock-salt lattice (lattice constant
$a=0.5$ nm) means that the electron is directly surrounded by gadolinium
cations. The distance to the first magnetic coordination shell (6
Gd ions) is $d_{1}=a/2=0.25$ nm. The second and third shells of magnetic
ions lie at $d_{2}=a\sqrt{3}/2\approx0.433$ nm (8 ions) and $d_{3}=a\sqrt{5}/2\approx0.559$
nm (24 ions), respectively. Owing to the exponential decay of the ground-state probability density, $|\varphi(r)|^{2}\propto\exp(-2r/a_{B})$, even in the extremely collapsed
state ($a_{B}=0.6$ nm) the electron still exhibits substantial overlap
with these shells: the probability density reaches approximately 0.43
 of its central value on the first shell, 0.23  on the second, and 0.15 on the third. 
Consequently, the localized electron continues to polarize strongly a collection of 38 magnetic Gd ions. By encompassing such a large number of spins, the polaron retains its collective character. This fully justifies the applicability of the thermodynamic variance approximation and the macroscopic Ginzburg–Landau–Wilson fluctuation theory, even in the small-polaron regime.

To fully visualize the breadth of this phenomenon across the parameter
space, we present three-dimensional surface maps of the bound magnetic
polaron's properties. Figure~\ref{fig:surface_Delta0} displays the
most probable internal spin splitting $\bar{\Delta}(J_{c},T)$. The
surface clearly reveals an extended topological region where a robust
local spin splitting ($\bar{\Delta}\gg0$) emerges spontaneously at
temperatures significantly above the bulk Curie temperature $T_{c}$,
bounded by the effective ordering temperature critical line $T^{*}(J_{c})$.

\begin{figure}[htbp]
\centering \includegraphics[width=0.8\textwidth]{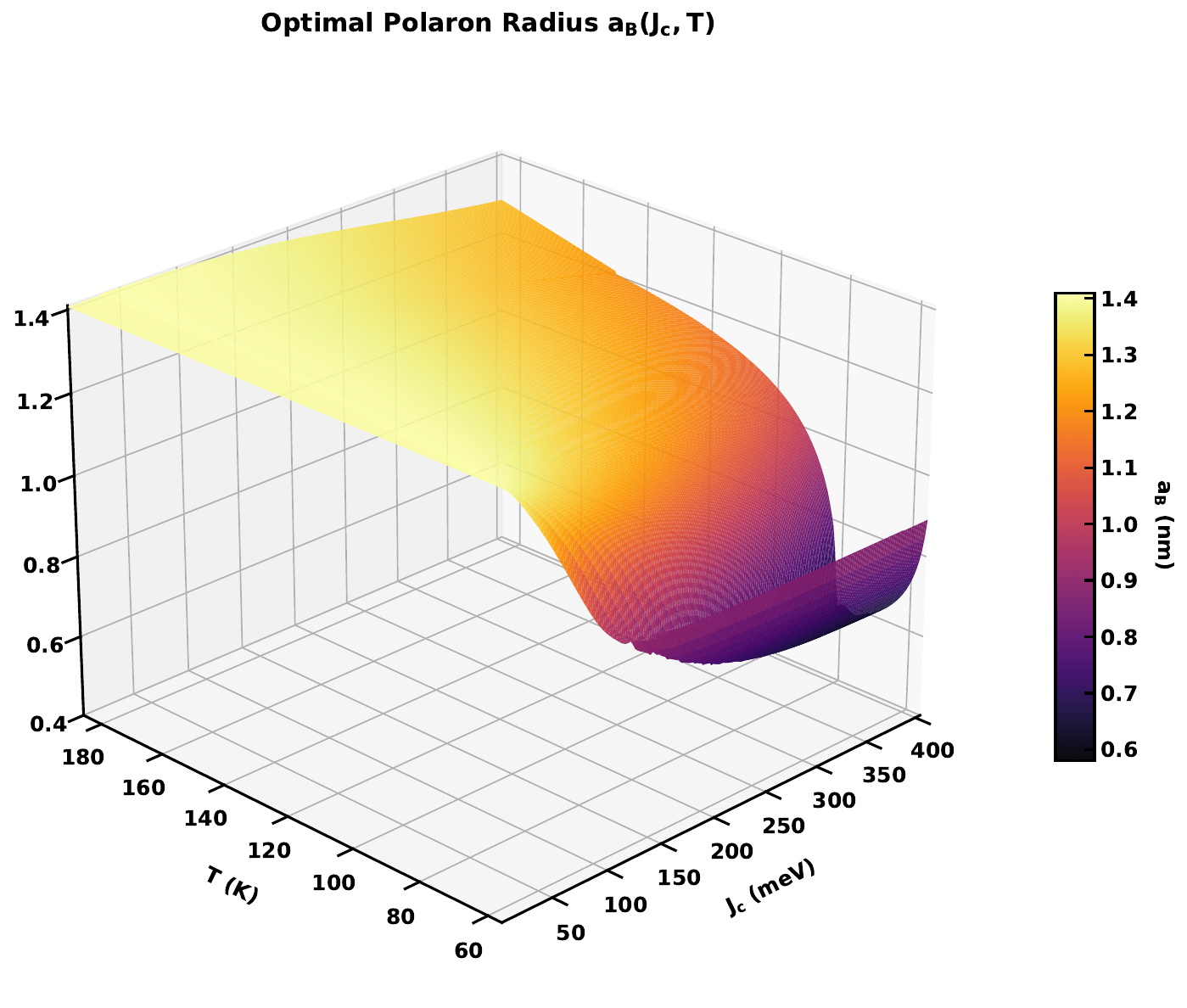} \caption{Three-dimensional surface plot of the optimal polaron radius $a_{B}(J_{c},T)$.
The deep valley indicates the region of strong magnetic self-trapping,
driven by the non-linear positive feedback between the kinetic energy
cost and the $1/a_{B}^{3}$ gain from non-Gaussian magnetization fluctuations.}
\label{fig:surface_aB} 
\end{figure}

Simultaneously, the manifestation of magnetic self-trapping is vividly
captured in Figure~\ref{fig:surface_aB}, which maps the optimal
Bohr radius $a_{B}(J_{c},T)$. The plot exposes a pronounced "valley"
(or cliff) of deep spatial localization. As the exchange interaction
$J_{c}$ intensifies, the onset of orbital contraction shifts to progressively
higher temperatures, directly mirroring the thermodynamic stabilization
provided by the anharmonic fluctuations. The sharp edge of this valley
corresponds to the polaron collapse discussed earlier.

To the best of our knowledge, the present work provides the first
rigorous microscopic model beyond the Gaussian approximation that
quantitatively predicts stable, ferromagnetically ordered bound magnetic
polarons with finite spontaneous spin splitting in the paramagnetic
phase of a ferromagnetic semiconductor. The results have direct implications
for the magnetic phase diagram of GdN (and related compounds EuO,
EuS etc.). At realistic doping levels the polaron density is high
enough for these ordered BMPs to overlap and percolate, offering a
natural mechanism for the persistence of macroscopic ferromagnetism
above $T_{c}$. This picture is fully consistent with experimental
observations of nitrogen-vacancy-mediated magnetism in GdN thin films
\cite{Natali2013PRB}.

\section{Dielectric Catastrophe and Magnetic Self-Trapping near the Metal-Insulator
Transition}

It is fundamentally instructive to examine how the static dielectric
constant $\varepsilon$ of the host material influences the stability
and spatial extent of bound magnetic polarons, particularly as the
system approaches the Mott metal--insulator transition (MIT). In
the vicinity of the critical donor concentration $n_{c}$, enhanced
screening leads to a dielectric catastrophe, causing the effective
static dielectric constant to diverge ($\varepsilon\to\infty$).

To rigorously capture this within our non-perturbative variational
framework, we must strictly distinguish the optimal (variational)
polaron radius $a_{{\rm opt}}$ from the bare hydrogenic Bohr radius
$a_{B0}(\varepsilon)$. The background dielectric constant defines
\emph{only} the bare initial state. The bare Bohr radius and the effective
Rydberg energy scale with the relative dielectric scaling parameter
$\tilde{\varepsilon}=\varepsilon/\varepsilon_{{\rm bulk}}$ as: 
\begin{equation}
a_{B0}(\tilde{\varepsilon})=a_{B0}^{{\rm bulk}}\cdot\tilde{\varepsilon},\qquad Ry^{*}(\tilde{\varepsilon})=\frac{Ry_{{\rm bulk}}^{*}}{\tilde{\varepsilon}^{2}}.
\end{equation}

Consequently, the single-particle Hamiltonian expectation value $\langle\varphi|\mathcal{H}_{1}|\varphi\rangle$,
comprising the kinetic energy cost of localization and the Coulomb
binding gain, becomes explicitly dependent on $\tilde{\varepsilon}$
via the variational parameter $a_{{\rm opt}}$: 
\begin{equation}
\langle\varphi|\mathcal{H}_{1}|\varphi\rangle=Ry^{*}(\tilde{\varepsilon})\left[\left(\frac{a_{B0}(\tilde{\varepsilon})}{a_{{\rm opt}}}\right)^{2}-2\frac{a_{B0}(\tilde{\varepsilon})}{a_{{\rm opt}}}\right]=\frac{\hbar^{2}}{2m^{*}a_{{\rm opt}}^{2}}-\frac{e^{2}}{4\pi\varepsilon_{0}\varepsilon a_{{\rm opt}}}.
\end{equation}

\begin{figure}[htbp]
\centering \includegraphics[width=0.8\textwidth]{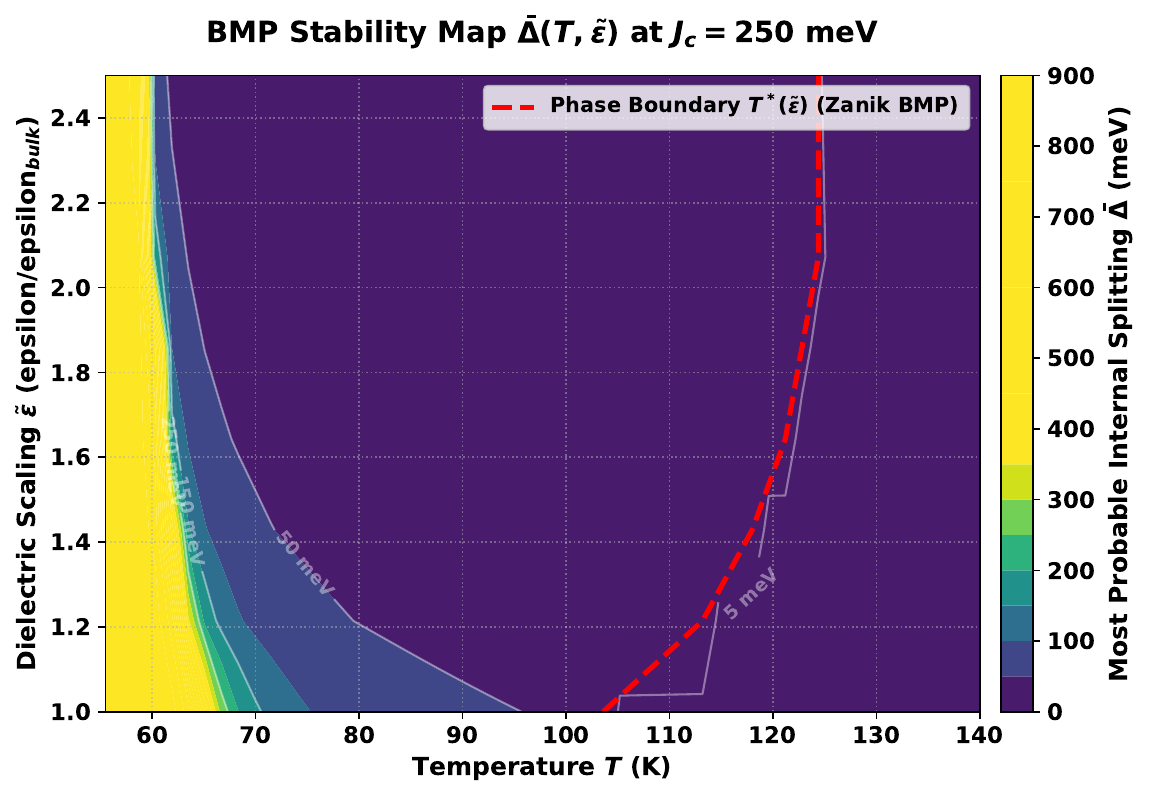}
\caption{Two-dimensional phase map defining the boundary of the bound magnetic
polaron stability region in the $(J_{c},T)$ plane. The contour marks
the effective polaron ordering temperature $T^{*}(J_{c})$.}
\label{fig:phase_map_2D} 
\end{figure}

A profound physical consequence emerges from this formulation. As
the system approaches the MIT and $\tilde{\varepsilon}$ diverges,
the Coulomb binding potential vanishes ($\propto1/\varepsilon$).
Without the magnetic exchange interaction, the donor electron would
completely delocalize ($a_{{\rm opt}}\to\infty$). However, the presence
of the localized $4f$ spins triggers intense \emph{magnetic self-trapping}.
To minimize the total free energy via the exchange interaction and
anharmonic thermodynamic fluctuations, the electron is forced to contract
its orbital. The interplay between this dielectric scaling and the
robustness of the ferromagnetic bubble is summarized in the macroscopic
phase map. As illustrated in Figure~\ref{fig:phase_map_2D}, the
parameter space defines an extended region of the $(J_{c},T)$ plane
in which stable, spontaneously magnetized BMPs ($\bar{\Delta}\neq0$)
exist well above the bulk Curie temperature. The contour of vanishing
$\bar{\Delta}$ defines an effective polaron critical line $T^{*}(J_{c})$
lying significantly above $T_{c}$ for all realistic coupling strengths.

Because the purely kinetic energy cost of this spatial collapse ($\propto1/a_{{\rm opt}}^{2}$)
is entirely independent of the background dielectric constant $\varepsilon$,
the polaron transitions from a state bound to the defect (Bound Magnetic
Polaron) to a practically free state (Free Magnetic Polaron). Its
localization is no longer maintained by the static impurity potential,
but rather self-sustained by the exchange-induced ferromagnetic bubble
it creates.

In this regime, the characteristic polaron energy $\varepsilon_{p}\propto1/a_{{\rm opt}}^{3}$
is determined by the self-consistent collapse of the wavefunction
rather than the vanishing Coulomb field. This reveals a critical window
near the MIT where macroscopic ferromagnetism is most robustly supported:
the initial dielectric softening unbinds the electrons from specific
defect sites, allowing them to self-trap magnetically, overlap, and
form a percolating ferromagnetic network well above the bulk Curie
temperature $T_{c}$.

\section{Summary and outlook}

We have extended the Dietl--Spałek theory of bound magnetic polarons
by rigorously incorporating the cubic and quartic anharmonic terms
of the Ginzburg--Landau--Wilson functional, going substantially
beyond both the original Gaussian treatment and our earlier renormalized-Gaussian
approach (which kept the donor radius fixed). Two complementary and
fully consistent generalizations have been developed: (i) a first-order
perturbative non-local correction valid for any finite spin correlation
length $\xi$, and (ii) a non-perturbative resummation of the dominant
local fluctuations into closed exponential form in the strict local
limit $\xi\to0$ that underlies the original framework. The validity
of the local resummation for macroscopic polarons is established by
a novel polaronic Ginzburg criterion (Appendix A), which demonstrates
that constraint-induced connected diagrams are suppressed by $\mathcal{O}(1/N_{{\rm eff}})$
with $N_{{\rm eff}}\gg1$, while the disconnected local thermodynamic
contributions resum exactly.

Applied to GdN ($T_{c}\approx55$ K), the theory eliminates the unphysical
divergence of the Gaussian model near $T_{c}$ and yields thermodynamically
stable BMPs carrying a finite spontaneous internal spin splitting
$\bar{\Delta}\neq0$ well into the paramagnetic phase. For realistic
exchange coupling $J_{c}=400$ meV the effective polaron ordering
temperature reaches $T^{*}\approx155$--$160$ K. Variational optimization
of the donor orbital reveals pronounced magnetic self-trapping, manifested
as strong orbital contraction and a characteristic kink in $a_{B}(T)$
at $T_{{\rm char}}\approx79$ K (above $T_{c}$), arising from the
competition between kinetic-energy cost and fluctuation-driven energy
gain. Importantly, despite the extreme spatial localization in the strong-coupling regime (down to $a_{B} \approx 0.6$ nm), the continuous-medium approximation remains robustly justified; the electron wave function still significantly overlaps with approximately 38 surrounding magnetic cations, firmly preserving the collective, macroscopic nature of the non-Gaussian thermodynamic fluctuations. The model further predicts an optimal donor-concentration window
near the Mott metal--insulator transition, where the dielectric catastrophe
enhances screening and thereby maximizes BMP stability and $T^{*}$.

These results establish the first quantitative microscopic mechanism
for persistent, BMP-mediated ferromagnetism above the bulk Curie temperature
via percolation of self-trapped ferromagnetic droplets. The findings
are consistent with nitrogen-vacancy-mediated magnetism observed in
GdN thin films and suggest several clear experimental signatures:
temperature-dependent optical spin splitting, anomalous magnetoresistance,
and magnetic susceptibility tails persisting above $T_{c}$. Future
work will incorporate finite-$\xi$ effects, explore the dielectric-tuned
doping window in greater detail, and extend the framework to related
compounds such as EuO, EuS, and magnetic oxides.

\section*{Acknowledgments}

The author utilized Gemini Pro and Grok (large language models by
Google and xAI, respectively) tools for language translation, manuscript
editing, verification of algebraic derivations, and assistance in
developing computational scripts. The author has rigorously reviewed,
validated, and edited all content generated or processed by the AI
and takes full responsibility for the accuracy, integrity, and final
presentation of the results.

\appendix

\section{Structure of mixed contributions and the polaronic Ginzburg criterion
for macroscopic resummation}

In the strictly local limit ($\xi\to0$), the exact evaluation of
the auxiliary functional $Z[\mathbf{J}]$ requires expanding the non-linear
operators and applying functional derivatives to the generated auxiliary
fields $\mathcal{F}_{i}(z)$ and $\mathcal{F}'_{i}(z)$. As defined
in the main text, the function $\mathcal{F}'_{i}(z)$ is independent
of the external field $J_{i}(r)$, thus identically yielding $\delta\mathcal{F}'_{i}(z_{2})/\delta J_{k}(z_{1})=0$.
The only non-vanishing functional derivative linking distinct spatial
points is the effective propagator: 
\begin{equation}
\mathcal{P}_{i}(z_{1},z_{2})\equiv\frac{\delta\mathcal{F}_{i}(z_{2})}{\delta J_{i}(z_{1})}\bigg|_{\xi\to0}=-\frac{1}{\Phi_{i}\bar{\mu}_{i}^{2}}|\varphi(z_{1})|^{2}|\varphi(z_{2})|^{2}.\label{eq:D-propagator}
\end{equation}
This separable, non-local term does not originate from the thermodynamic
spin correlations of the host material (which vanish identically for
$\xi\to0$). Instead, it is a mathematical consequence of the global
constraint imposed by the Dirac delta function defining the macroscopic
spin splitting (Eq.~(\ref{eq:P-Delta}) of the main text).

\subsection{Operator definitions and second-order contribution}

Introduce the functional derivative operator $\hat{d}_{i,\mathbf{z}}\equiv\frac{\delta}{\delta J_{i}(\mathbf{z})}$.
The cubic and quartic parts of the vertex operator are 
\begin{align}
\hat{\mathcal{K}}_{\mathbf{z}} & =\hat{d}_{3,\mathbf{z}}^{3}+\hat{d}_{3,\mathbf{z}}\hat{d}_{1,\mathbf{z}}^{2}+\hat{d}_{3,\mathbf{z}}\hat{d}_{2,\mathbf{z}}^{2},\\
\hat{\mathcal{Q}}_{\mathbf{z}} & =\frac{1}{4}\left(\sum_{i=1}^{3}\hat{d}_{i,\mathbf{z}}^{4}+2\sum_{j<i}\hat{d}_{i,\mathbf{z}}^{2}\hat{d}_{j,\mathbf{z}}^{2}\right),
\end{align}
so that the full vertex operator reads $\hat{\mathcal{V}}_{\mathbf{z}}=M_{0}\hat{\mathcal{K}}_{\mathbf{z}}+\hat{\mathcal{Q}}_{\mathbf{z}}$.

The second-order contribution to the auxiliary functional is 
\begin{equation}
Z^{(2)}=\frac{(-\lambda)^{2}}{2!}\int d^{3}z_{1}\int d^{3}z_{2}\,\hat{\mathcal{V}}_{\mathbf{z}_{1}}\hat{\mathcal{V}}_{\mathbf{z}_{2}}\,\Xi[\mathbf{J}].
\end{equation}

When the first vertex $\hat{\mathcal{V}}_{\mathbf{z}_{2}}$ acts on
the Gaussian generating functional $\Xi$, it produces a polynomial
$\mathcal{W}_{\mathbf{z}_{2}}\Xi$, where $\mathcal{W}_{\mathbf{z}_{2}}$
contains combinations of $\mathcal{F}_{i}(z_{2})$ and $\mathcal{F}'_{i}(z_{2})$
(the explicit form for $n=1$ appears in Eq.~(\ref{eq:Z-exp-tilded})).
The second vertex $\hat{\mathcal{V}}_{\mathbf{z}_{1}}$ then acts
on this result: $\hat{\mathcal{V}}_{\mathbf{z}_{1}}(\mathcal{W}_{\mathbf{z}_{2}}\Xi)$.

By the Leibniz (product) rule for functional derivatives, this action
splits into two classes of terms.

\subsection{Disconnected (local thermodynamic) contributions}

These arise when all functional derivatives in $\hat{\mathcal{V}}_{\mathbf{z}_{1}}$
act directly on $\Xi$ (bypassing $\mathcal{W}_{\mathbf{z}_{2}}$).
They produce simple products of the local structures generated independently
at $\mathbf{z}_{1}$ and $\mathbf{z}_{2}$: 
\begin{equation}
\mathcal{W}_{\mathbf{z}_{1}}\mathcal{W}_{\mathbf{z}_{2}}\Xi.
\end{equation}
These are the ``purely local'' terms that survive in the exponential
resummation of (Eq.~(\ref{eq:Z-resum-local-tilded})) in the main
text. They correspond to independent thermodynamic fluctuations at
distant points and form the backbone of the non-perturbative local
theory. In the approximate expansion used in the main text (second
line of Eq.~(\ref{eq:Z-exp-tilded})), only these disconnected terms
are retained, leading to the product form that exponentiates exactly
in the local limit.

\subsection{Connected (constraint-induced mixed) contributions}

When any derivative from $\hat{\mathcal{V}}_{\mathbf{z}_{1}}$ strikes
the polynomial $\mathcal{W}_{\mathbf{z}_{2}}$ generated at $\mathbf{z}_{2}$,
it inserts the non-local propagator $\mathcal{P}_{i}(z_{1},z_{2})$.
These terms are ``connected'' by one or more propagators.

To illustrate the full structure explicitly, consider the pure longitudinal
cubic-cubic contribution $M_{0}^{2}\hat{d}_{3,\mathbf{z}_{1}}^{3}\hat{d}_{3,\mathbf{z}_{2}}^{3}\Xi$
(one representative channel; all other combinations follow analogously).

After acting with $\hat{d}_{3,\mathbf{z}_{2}}^{3}$ on $\Xi$ we have
\begin{equation}
\hat{d}_{3,\mathbf{z}_{2}}^{3}\Xi=\left(\mathcal{F}_{3,\mathbf{z}_{2}}^{3}+3\mathcal{F}_{3,\mathbf{z}_{2}}\mathcal{F}'_{3,\mathbf{z}_{2}}\right)\Xi\equiv A_{2}\Xi.
\end{equation}

Applying the three derivatives at $\mathbf{z}_{1}$ step-by-step (using
the product rule repeatedly) yields four distinct topological classes
classified by the number of $\mathcal{P}_{12}\equiv\mathcal{P}_{3}(z_{1},z_{2})$
insertions (i.e., the number of Wick contractions between the two
vertices):

1. \textit{Zero contractions (fully disconnected)}: 
\begin{equation}
\left(\mathcal{F}_{3,\mathbf{z}_{1}}^{3}+3\mathcal{F}_{3,\mathbf{z}_{1}}\mathcal{F}'_{3,\mathbf{z}_{1}}\right)\left(\mathcal{F}_{3,\mathbf{z}_{2}}^{3}+3\mathcal{F}_{3,\mathbf{z}_{2}}\mathcal{F}'_{3,\mathbf{z}_{2}}\right)\Xi
\end{equation}

2. \textit{Single contraction (one} $\mathcal{P}_{12}$\textit{):}
\begin{equation}
9\mathcal{P}_{12}\left(\mathcal{F}_{3,\mathbf{z}_{1}}^{2}+\mathcal{F}'_{3,\mathbf{z}_{1}}\right)\left(\mathcal{F}_{3,\mathbf{z}_{2}}^{2}+\mathcal{F}'_{3,\mathbf{z}_{2}}\right)\Xi
\end{equation}

3. \textit{Double contraction (two} $\mathcal{P}_{12}^{2}$\textit{):
} 
\begin{equation}
18\mathcal{P}_{12}^{2}\mathcal{F}_{3,\mathbf{z}_{1}}\mathcal{F}_{3,\mathbf{z}_{2}}\Xi
\end{equation}

4. \textit{Triple contraction (three} $\mathcal{P}_{12}^{3}$\textit{):}
\begin{equation}
6\mathcal{P}_{12}^{3}\Xi
\end{equation}

(The numerical prefactors are the standard combinatorial weights from
Wick's theorem for contracting three derivatives at each vertex.)

The connected terms (classes 2--4) involve one or more factors of
the propagator $\mathcal{P}_{12}\propto|\varphi(z_{1})|^{2}|\varphi(z_{2})|^{2}$,
which is precisely the separable kernel responsible for the global
constraint. All higher-order terms ($n>2$) and all mixed (cubic-quartic,
quartic-quartic) contributions generate analogous connected diagrams
with multiple propagators.

In the local approximation, the disconnected contributions dominate
the physics of independent thermodynamic fluctuations and can be resummed
exactly into the exponential form. The connected contributions, however,
are systematically suppressed by powers of $1/N_{{\rm eff}}$ (see
scaling analysis below) and can be neglected for macroscopic polarons.

\subsection{Inductive structure of higher-order terms}

To prove that the local resummation of the disconnected terms remains
valid to \emph{all} orders in the anharmonic expansion, we employ
mathematical induction on the perturbation order $n$.

\textbf{Base cases:} 
\begin{itemize}
\item $n=0$: The Gaussian generating functional $\Xi[\mathbf{J}]$ contains
only local structures (after $\xi\to0$ regularization); no vertices,
hence no connected terms. 
\item $n=1$: A single application of $\hat{\mathcal{V}}_{\mathbf{z}}$
produces purely local polynomials in $\mathcal{F}_{i}(\mathbf{z})$
and $\mathcal{F}'_{i}(\mathbf{z})$ (no inter-vertex contractions
possible). The entire $Z^{(1)}$ is disconnected. 
\end{itemize}
\textbf{Induction hypothesis:} Assume that at order $n$, 
\begin{equation}
Z^{(n)}=D^{(n)}+C^{(n)},
\end{equation}
where $D^{(n)}$ is the sum of fully disconnected products of $n$
independent local vertex factors (polynomials in $\mathcal{F}_{i}(\mathbf{z}_{k}),\mathcal{F}'_{i}(\mathbf{z}_{k})$
at distinct points $\mathbf{z}_{k}$), and $C^{(n)}$ collects all
terms containing at least one inter-vertex contraction via the propagator
$\mathcal{P}_{i}(z_{j},z_{k})$, each suppressed by $\mathcal{O}(1/N_{{\rm eff}})$.

\textbf{Inductive step ($n\to n+1$)}: We obtain $Z^{(n+1)}$ by applying
$\hat{\mathcal{V}}_{\mathbf{z}_{n+1}}$ to $Z^{(n)}=D^{(n)}+C^{(n)}$.
By the Leibniz (product) rule: 
\begin{itemize}
\item Terms where $\hat{\mathcal{V}}_{\mathbf{z}_{n+1}}$ acts solely on
the Gaussian $\Xi$ (bypassing previous polynomials) produce the fully
disconnected product $D^{(n+1)}$ - exactly the next term in the exponential
resummation. 
\item Terms where any derivative strikes a polynomial $W$ generated by
the previous $n$ vertices necessarily insert at least one factor
of $\mathcal{P}_{i}(z_{n+1},z_{j})$ ($j\le n$). Crucially, because
the internal propagator $\mathcal{P}_{i}(z_{n+1},z_{j})$ is governed
by the separable envelope kernel $\propto|\varphi(z_{n+1})|^{2}|\varphi(z_{j})|^{2}$,
the required spatial integration over the new vertex coordinate $\mathbf{z}_{n+1}$
explicitly extracts a factor of the local constraint constant $c_{0}\sim1/V_{p}\propto1/N_{{\rm eff}}$.
Thus, inheriting the suppression from $C^{(n)}$, every new internal
connection intrinsically introduces an extra volume-suppression factor
$\mathcal{O}(1/N_{{\rm eff}})$. 
\end{itemize}
Each internal Wick contraction contributes an extra factor $c_{0}\sim1/V_{p}\propto1/N_{{\rm eff}}$
(from the separable kernel $\mathcal{P}\propto|\varphi(z_{1})|^{2}|\varphi(z_{2})|^{2}$).
Therefore the relative weight of all connected contributions remains
$\mathcal{O}(1/N_{{\rm eff}})$ (or smaller) uniformly in $n$.

By induction the statement holds for every finite order. Consequently
the infinite series of purely local (disconnected) terms can be resummed
exactly into the closed exponential form of Eq.~(\ref{eq:Z-resum-local-tilded})
of the main text, while the constraint-induced connected corrections
are negligible for macroscopic polarons ($N_{{\rm eff}}\gg1$).

\subsection{Explicit second-order calculation in the paramagnetic regime
($M_0=0$, $\Delta_0=0$)}

In the paramagnetic phase of primary interest for the stability of BMPs
above $T_c$ (where $M_0=0$ and $\Delta_0=0$), the cubic vertices
vanish and the vertex operator reduces to the pure quartic form
\begin{equation}
\hat{\mathcal{Q}}_{\mathbf{z}}=\frac{1}{4}\left(\sum_{i=1}^{3}\hat{d}_{i,\mathbf{z}}^{4}+2\sum_{j<i}\hat{d}_{i,\mathbf{z}}^{2}\hat{d}_{j,\mathbf{z}}^{2}\right).
\end{equation}
The second-order contribution then reads
\begin{equation}
Z^{(2)}=\frac{\tilde{\lambda}^{2}}{2!}\iint d^{3}z_{1}\,d^{3}z_{2}\,
\hat{\mathcal{Q}}_{\mathbf{z}_{1}}\hat{\mathcal{Q}}_{\mathbf{z}_{2}}\,\Xi[\tilde{\mathbf{J}}].
\end{equation}
Acting with the operator at $\mathbf{z}_{2}$ produces the local polynomial
$\mathcal{W}^{(4)}_{\mathbf{z}_{2}}\Xi$, whose explicit form (from the
$n=1$ term in the main-text expansion) is
\begin{equation}
\mathcal{W}^{(4)}_{\mathbf{z}_{2}}=\frac{1}{4}\Biggl[\sum_{i=1}^{3}\bigl(3(\mathcal{F}'_{i,2})^{2}+6\mathcal{F}'_{i,2}\mathcal{F}_{i,2}^{2}+\mathcal{F}_{i,2}^{4}\bigr)
+2\sum_{j<i}\bigl(\mathcal{F}_{j,2}^{2}+\mathcal{F}'_{j,2}\bigr)\bigl(\mathcal{F}_{i,2}^{2}+\mathcal{F}'_{i,2}\bigr)\Biggr],
\end{equation}
with the shorthand $\mathcal{F}_{i,2}\equiv\mathcal{F}_{i}(\mathbf{z}_{2})$ etc.

To obtain $Z^{(2)}$ we must apply $\hat{\mathcal{Q}}_{\mathbf{z}_{1}}$ to
$\mathcal{W}^{(4)}_{\mathbf{z}_{2}}\Xi$. By the Leibniz rule this again
splits into disconnected (local) products $\mathcal{W}^{(4)}_{\mathbf{z}_{1}}
\mathcal{W}^{(4)}_{\mathbf{z}_{2}}\Xi$ that are resummed into the
exponential factor, and connected contributions that insert one or more
factors of the constraint propagator
\begin{equation}
\mathcal{P}_{i}(z_1,z_2)=-\frac{1}{\Phi_i\bar{\mu}_i^2}|\varphi(z_1)|^2|\varphi(z_2)|^2
\end{equation}
(note that $\delta\mathcal{F}'_i/\delta\tilde{J}_k=0$, so only $\mathcal{F}$
fields generate non-local links).

As a concrete illustration we evaluate the dominant diagonal longitudinal
channel $\frac{1}{16}\hat{d}_{3,\mathbf{z}_1}^4\hat{d}_{3,\mathbf{z}_2}^4\Xi$.
After the action at $\mathbf{z}_2$ one obtains the polynomial
$\bigl(\mathcal{F}_{3,2}^4+6\mathcal{F}'_{3,2}\mathcal{F}_{3,2}^2+3(\mathcal{F}'_{3,2})^2\bigr)\Xi$.
Applying the four derivatives at $\mathbf{z}_1$ and collecting terms according to the number $k$ of Wick contractions yields five topological classes ($k=0$ to $k=4$).
The case $k=0$ (zero contractions) is completely disconnected. This is precisely the term that survives in the exponential resummation.
The four **connected** classes ($k \ge 1$) are:  

\begin{enumerate}
\item Single contraction ($k=1$): 
$16\mathcal{P}_{12}\bigl(\mathcal{F}_{3,1}^3+3\mathcal{F}_{3,1}\mathcal{F}'_{3,1}\bigr)
\bigl(\mathcal{F}_{3,2}^3+3\mathcal{F}_{3,2}\mathcal{F}'_{3,2}\bigr)\Xi$
\item Double contraction ($k=2$): 
$72\mathcal{P}_{12}^2\bigl(\mathcal{F}_{3,1}^2+\mathcal{F}'_{3,1}\bigr)
\bigl(\mathcal{F}_{3,2}^2+\mathcal{F}'_{3,2}\bigr)\Xi$
\item Triple contraction ($k=3$): 
$96\mathcal{P}_{12}^3\mathcal{F}_{3,1}\mathcal{F}_{3,2}\Xi$
\item Quadruple contraction ($k=4$): 
$24\mathcal{P}_{12}^4\Xi$
\end{enumerate}

(The prefactors are the standard combinatorial weights from Wick's
theorem.) Classes 1--4 are connected by the global constraint and will be shown below to
be suppressed by additional powers of $1/N_{\rm eff}$.

This explicit combinatorics for the pure-quartics paramagnetic case
confirms the general structure already visible in the cubic-cubic example
of the preceding subsection and justifies retaining only the disconnected
terms in the non-perturbative local resummation.

\subsection{Short remark on the connection to $\phi^4$ field theory}

The combinatorial structure generated by repeated functional differentiation
of the Gaussian generating functional is formally identical to the
Feynman-diagram expansion of a classical $\phi^4$ scalar field theory
(with the magnetization fluctuations playing the role of the order-parameter
field). However, because we work in the strict local limit $\xi\to0$
and exploit the macroscopic size of the polaron ($N_{\rm eff}\gg1$),
we can resum the dominant disconnected contributions exactly into
exponential form without invoking the full apparatus of renormalization
or the renormalization group. The connected (constraint-induced)
diagrams remain perturbatively small, as quantified by the polaronic
Ginzburg criterion derived in the scaling analysis below. The first-order
non-local correction of Sec.~II.D corresponds to the leading connected
Yukawa-exchange diagrams of the underlying $\phi^4$ theory when a finite
correlation length is restored.

\subsection{Scaling relations with BMP volume}

To rigorously justify the exponential resummation of the local thermodynamic
terms, we perform a detailed scaling analysis with respect to the
effective polaron volume $V_{p}\propto a_{B}^{3}$. The probability
density of the macroscopic hydrogenic envelope satisfies $|\varphi(\mathbf{r})|^{2}\sim1/V_{p}$.
Consequently, the bare spatial overlap integrals scale as

\begin{equation}
\overline{I}_{n}\equiv\int d^{3}r\,|\varphi(\mathbf{r})|^{2n}\sim V_{p}\left(\frac{1}{V_{p}}\right)^{n}=V_{p}^{1-n}.
\end{equation}

Specifically, $\overline{I}_{2}\sim V_{p}^{-1}$, $\overline{I}_{3}\sim V_{p}^{-2}$,
and $\overline{I}_{4}\sim V_{p}^{-3}$. The local propagator constraint
constant, $c_{0}\equiv\frac{1}{2}\left(\frac{\alpha}{g\mu_{B}}\right)^{2}\overline{I}_{2}$,
which governs the variance of the Gaussian fluctuations, therefore
scales directly with the inverse volume:

\begin{equation}
c_{0}\sim V_{p}^{-1}.
\end{equation}

As established in the main text, the effective vertex integrals inherently
absorb the constraint constant generated by the functional derivatives.
Strikingly, these effective integrals scale exactly linearly with
the polaron volume:

\begin{equation}
\mathcal{I}_{n}\equiv\frac{\overline{I}_{n}}{c_{0}^{n}}\Bigl(\frac{\alpha}{g\mu_{B}}\Bigr)^{n}\sim\frac{V_{p}^{1-n}}{(V_{p}^{-1})^{n}}(V_{p}^{0})^{n}=V_{p}^{1}.
\end{equation}

Finally, in the physically relevant thermodynamic regime, the typical
squared amplitude of the spin-splitting fluctuations $\Lambda^{2}$
is of the order of the Gaussian variance, meaning $\Lambda^{2}\sim c_{0}\sim V_{p}^{-1}$.

\subsection{Scaling for disconnected (purely local thermodynamic) contributions}

The disconnected contributions form the backbone of the purely local
thermodynamic fluctuations. The leading local anharmonic term in the
paramagnetic phase ($M_{0}=0$) arises from the quartic vertex. Using
the effective vertex integrals, its energy contribution in the exponent
reads:

\begin{equation}
E_{{\rm local}}^{(4)}\propto\frac{\lambda}{k_{B}T}\mathcal{I}_{4}\Lambda^{4}.
\end{equation}

Substituting the derived scaling relations $\mathcal{I}_{4}\sim V_{p}^{1}$
and $\Lambda^{4}\sim(V_{p}^{-1})^{2}=V_{p}^{-2}$, we obtain:

\begin{equation}
E_{{\rm local}}^{(4)}\sim(V_{p}^{0})\cdot(V_{p}^{1})\cdot(V_{p}^{-2})=V_{p}^{-1}.
\end{equation}

Thus, the purely thermodynamic contribution vanishes exactly as $1/V_{p}$
(or equivalently $1/N_{{\rm eff}}$) for macroscopic polarons, ensuring
that the local anharmonic corrections remain finite and thermodynamically
stable.

\subsection{Scaling for connected (constraint-induced) mixed contributions}

The connected terms arise from multiple functional derivatives acting
across different spatial points, intrinsically linking them via the
non-local constraint propagator $\mathcal{P}_{12}\propto c_{0}$.
The leading mixed term at second order in $\lambda$ originates from
a single Wick contraction between two quartic vertices:

\begin{equation}
E_{{\rm mix}}\propto\left(\frac{\lambda}{k_{B}T}\right)^{2}\Lambda^{6}(\mathcal{I}_{4})^{2}c_{0}.
\end{equation}

Notice how the explicit reliance on the effective vertex integrals
$\mathcal{I}_{4}$ cleanly isolates the single residual factor of
$c_{0}$ originating from the internal contraction. Inserting the
scaling relations yields:

\begin{equation}
E_{{\rm mix}}\sim(V_{p}^{0})\cdot(V_{p}^{-3})\cdot(V_{p}^{2})\cdot(V_{p}^{-1})=V_{p}^{-2}.
\end{equation}

Analogous scaling holds for all higher-order connected diagrams and
for the ferromagnetic phase ($M_{0}\neq0$), where cubic vertices
generate mixed terms containing combinations of $\mathcal{I}_{3}$
and $\mathcal{I}_{4}$. In every case, each internal Wick contraction
introduces an additional factor of $c_{0}\sim V_{p}^{-1}$.

\subsection{Polaronic Ginzburg criterion}

The ratio of the leading constraint-induced (connected) contribution
to the dominant thermodynamic (disconnected) term is 
\begin{equation}
\frac{E_{{\rm mix}}}{E_{{\rm local}}}\propto\Lambda^{2}\mathcal{I}_{4}c_{0}\sim V_{p}^{-1}\propto\frac{1}{N_{{\rm eff}}}.\label{eq:Ginzburg-polaronic}
\end{equation}
For realistic donor radii in ferromagnetic semiconductors ($a_{B}\sim1$--$2$,nm),
one finds $N_{{\rm eff}}\sim10^{2}$--$10^{3}$. In this macroscopic
polaron limit, the non-local mixed terms are strongly suppressed.
The exponential resummation of the purely local terms therefore becomes
non-perturbatively exact within the local ($\xi\to0$) approximation
inherent to the original Dietl--Spałek framework.

This polaronic Ginzburg criterion is the direct analogue of the classic
Ginzburg--Levanyuk criterion, but here the control parameter is the
macroscopic effective volume of the bound magnetic polaron rather
than the correlation volume near criticality. While the standard criterion
signals the breakdown of the Gaussian approximation when the mean-square
fluctuation of the order parameter within a correlation volume becomes
comparable to the square of its mean value, here the same physical
role is played by the strong suppression ($\sim1/N_{{\rm eff}}$)
of the constraint-induced connected diagrams relative to the dominant
purely local thermodynamic fluctuations. Because realistic bound magnetic
polarons enclose a macroscopic number of spins ($N_{{\rm eff}}\sim10^{2}$--$10^{3}$),
the local resummation of the anharmonic terms becomes non-perturbatively
accurate.

To rigorously quantify the validity of neglecting the constraint-induced connected diagrams, we evaluated their relative contribution in the second order of perturbation theory. The connected terms are subjected to a powerful double suppression: topologically by the macroscopic effective volume $\mathcal{O}(1/N_{\text{eff}})$ and parametrically by the squared reduced anharmonicity coupling $\lambda_p^2 \propto \chi^4 / a_B^6$. In the physically crucial regime of magnetic self-trapping ($T_{\text{char}} \approx 79$ K), the host susceptibility is sufficiently small that the relative weight of the connected diagrams is strictly negligible ($\sim 10^{-9}$). 

\begin{figure}[htbp]
\centering 
\includegraphics[width=0.75\textwidth]{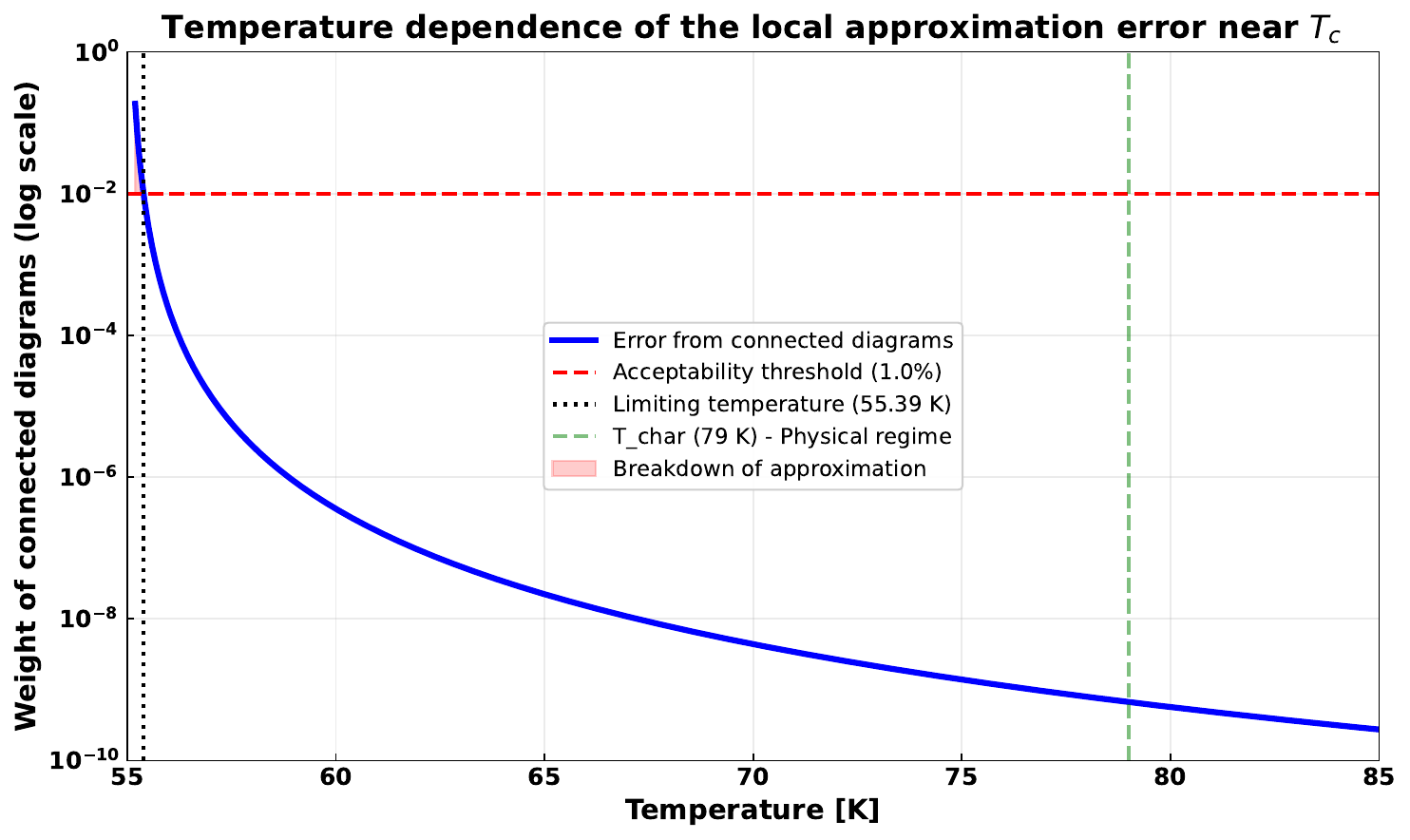}
\caption{Temperature dependence of the relative weight of the constraint-induced connected diagrams evaluated in the second order of the anharmonic expansion. The curve represents the quantitative error of the non-perturbative local approximation ($N_{\text{eff}}=38$, $a_B=0.6$ nm). The logarithmic scale reveals that in the physically relevant regime of magnetic self-trapping ($T_{\text{char}} \approx 79$ K), the error is strictly negligible ($\sim 10^{-9}$). The approximation remains robust (error $<1\%$) down to a critical boundary of $T \approx 55.4$ K, safely below the onset of the structural crossover.}
\label{fig:Ginzburg_error} 
\end{figure}

Crucially, this stability is maintained even as the system approaches the macroscopic phase transition ($T \to T_c^+$) and $\chi$ diverges. As the temperature is lowered towards $T_c$, the optimal polaron orbit undergoes spatial relaxation, expanding to $a_B \approx 0.9$ nm at $T = 56$ K. This expansion significantly increases the effective number of enclosed spins ($N_{\text{eff}} > 115$), dynamically buffering the growth of the parametric coupling with enhanced volumetric suppression. Numerical analysis confirms that the cumulative weight of the connected diagrams remains below $1\%$ down to $T \approx 55.4$ K, as shown in Figure~\ref{fig:Ginzburg_error}. Therefore, the non-perturbative resummation of purely local thermodynamic fluctuations is physically robust and mathematically justified across the entire paramagnetic regime relevant to BMP formation. It is worth noting that the bare exchange coupling constant $J_c$ analytically cancels out in the ratio of connected to disconnected diagrams. Consequently, the relative error presented in Fig. 8 depends exclusively on the topological parameters of the self-trapped state ($a_B = 0.6$ nm, $N_{\text{eff}} = 38$) and the thermodynamic susceptibility of the host lattice, making this error bound universal for any $J_c$ that drives the system into this localization regime.

\section{Analytical properties of the spin splitting distribution: angular
integration and limiting cases}

This appendix provides the detailed algebraic derivation of the angular
integration that leads to the probability distribution $P(\boldsymbol{\Delta})$
in the local limit $\xi\to0$. The starting point is the resummed
auxiliary functional $Z[\mathbf{J}=0]$ given by Eq.~(\ref{eq:Z-resum-local-tilded})
of the main text. After performing the change of variables to the
modulus $\Delta$ and the angle $\theta$ between $\boldsymbol{\Delta}$
and the uniform field $\boldsymbol{\Delta}_{0}$, the exponent in
the integrand becomes the quadratic polynomial 
\begin{equation}
E(x)=Ax^{2}+Bx+C,
\end{equation}
where $x=\cos\theta$. The explicit coefficients, obtained by collecting
the Gaussian, cubic ($\propto M_{0}$) and quartic contributions,
read 
\begin{align}
A & =\frac{\Delta^{2}}{4}\left(\frac{1}{\Phi_{\perp}}-\frac{1}{\Phi_{\parallel}}\right)+\frac{\lambda\Delta^{2}}{16k_{B}T}\left[4M_{0}\mathcal{I}_{3}\Delta_{0}-\mathcal{I}_{4}(\Delta_{0}^{2}-2\Phi_{\parallel}+2\Phi_{\perp})\right],\label{eq:A}\\
B & =\frac{\Delta\Delta_{0}}{2\Phi_{\parallel}}-\frac{\lambda M_{0}\mathcal{I}_{3}\Delta}{8k_{B}T}\left[\Delta^{2}+3\Delta_{0}^{2}-2(3\Phi_{\parallel}+2\Phi_{\perp})\right]\nonumber \\
 & \quad+\frac{\lambda\mathcal{I}_{4}\Delta\Delta_{0}}{16k_{B}T}\left[\Delta^{2}+\Delta_{0}^{2}-2(3\Phi_{\parallel}+2\Phi_{\perp})\right],\label{eq:B}\\
C & =-\frac{\Delta_{0}^{2}}{4\Phi_{\parallel}}-\frac{\Delta^{2}}{4\Phi_{\perp}}+\frac{\lambda M_{0}\mathcal{I}_{3}\Delta_{0}}{8k_{B}T}\left[\Delta^{2}+\Delta_{0}^{2}-2(3\Phi_{\parallel}+2\Phi_{\perp})\right]\nonumber \\
 & \quad-\frac{\lambda\mathcal{I}_{4}}{4k_{B}T}\left[\frac{1}{16}(\Delta^{2}+\Delta_{0}^{2})^{2}-\frac{1}{4}\Delta_{0}^{2}(3\Phi_{\parallel}+2\Phi_{\perp})-\frac{1}{4}\Delta^{2}(\Phi_{\parallel}+4\Phi_{\perp})+\frac{3}{4}\Phi_{\parallel}^{2}+\Phi_{\parallel}\Phi_{\perp}+2\Phi_{\perp}^{2}\right].\label{eq:C}
\end{align}
(The normalization integrals $\mathcal{I}_{n}$ are defined in the
main text.)

Because the exponent is independent of the azimuthal angle $\phi$,
the angular integral reduces to 
\begin{equation}
I_{{\rm ang}}=2\pi\int_{-1}^{1}\exp(Ax^{2}+Bx+C)\,dx.
\end{equation}
Completing the square, 
\begin{equation}
Ax^{2}+Bx+C=A\left(x+\frac{B}{2A}\right)^{2}-\frac{B^{2}}{4A}+C,
\end{equation}
the integral over the finite interval $[-1,1]$ can be expressed exactly
with the error function $\operatorname{erf}$ or the imaginary error
function $\operatorname{erfi}$.

\subsubsection{Case $A<0$}

Let $\alpha=\sqrt{-A}$ and $\gamma=B/(2\alpha)$. Then 
\begin{equation}
I_{{\rm ang}}=\pi\sqrt{\frac{\pi}{-A}}\exp\left(C-\frac{B^{2}}{4A}\right)\Bigl[\operatorname{erf}(\alpha-\gamma)+\operatorname{erf}(\alpha+\gamma)\Bigr].
\end{equation}
Equivalently (without auxiliary symbols): 
\begin{equation}
I_{{\rm ang}}=\pi\sqrt{\frac{\pi}{-A}}\exp\left(C-\frac{B^{2}}{4A}\right)\Bigl[\operatorname{erf}\Bigl(\sqrt{-A}\bigl(1+\tfrac{B}{2A}\bigr)\Bigr)-\operatorname{erf}\Bigl(\sqrt{-A}\bigl(-1+\tfrac{B}{2A}\bigr)\Bigr)\Bigr].
\end{equation}

\subsubsection{Case $A>0$}

Let $\alpha=\sqrt{A}$ and $\gamma=B/(2\alpha)$. Then 
\begin{equation}
I_{{\rm ang}}=\pi\sqrt{\frac{\pi}{A}}\exp\left(C-\frac{B^{2}}{4A}\right)\Bigl[\operatorname{erfi}(\alpha-\gamma)+\operatorname{erfi}(\alpha+\gamma)\Bigr].
\end{equation}
Equivalently: 
\begin{equation}
I_{{\rm ang}}=\pi\sqrt{\frac{\pi}{A}}\exp\left(C-\frac{B^{2}}{4A}\right)\Bigl[\operatorname{erfi}\Bigl(\sqrt{A}\bigl(1+\tfrac{B}{2A}\bigr)\Bigr)-\operatorname{erfi}\Bigl(\sqrt{A}\bigl(-1+\tfrac{B}{2A}\bigr)\Bigr)\Bigr].
\end{equation}

\subsubsection{The Gaussian Limit with Finite Magnetization ($\lambda=0$, $\Delta_{0}\protect\neq0$)}

In the purely Gaussian limit ($\lambda=0$), the anisotropic thermodynamic
fluctuations still generate a non-vanishing curvature coefficient
$A=\frac{\Delta^{2}}{4}(\Phi_{\perp}^{-1}-\Phi_{\parallel}^{-1})$,
meaning the exact angular integration retains the error-function form.
However, if one assumes an isotropic fluctuation environment ($\Phi_{\parallel}=\Phi_{\perp}\equiv\Phi$),
the curvature vanishes ($A=0$), and the linear coefficients simplify
to: 
\begin{equation}
B=\frac{\Delta\Delta_{0}}{2\Phi},\quad C=-\frac{\Delta^{2}+\Delta_{0}^{2}}{4\Phi}.
\end{equation}
The angular integral $I_{{\rm ang}}$ then recovers exactly the form
known from the original DS theory: 
\begin{equation}
I_{ang}(\Delta;\Delta_{0})=4\pi\exp\left(-\frac{\Delta^{2}+\Delta_{0}^{2}}{4\Phi}\right)\frac{\sinh\left(\frac{\Delta\Delta_{0}}{2\Phi}\right)}{\frac{\Delta\Delta_{0}}{2\Phi}}.
\end{equation}
The above expression describes the influence of thermal magnetization
fluctuations on the spin splitting in the presence of an axis distinguished
by macroscopic magnetization. It is worth noting that despite the
lack of fourth-order stabilization, this formula is mathematically
correct as long as the magnetic susceptibility (contained in $\Phi$)
remains finite.

\subsubsection{Paramagnetic phase ($M_{0}=0$, $\Delta_{0}=0$)}

In the symmetric paramagnetic phase the curvature and linear coefficients
vanish identically ($A=B=0$). The exponent reduces to the purely
$\Delta$-dependent term 
\begin{equation}
C(\Delta)=-\frac{\Delta^{2}}{4\Phi}-\frac{\lambda\mathcal{I}_{4}}{4k_{B}T}\left(\frac{1}{16}\Delta^{4}-\frac{5}{4}\Phi\Delta^{2}+\frac{15}{4}\Phi^{2}\right).
\end{equation}
Consequently the angular integral is trivial, $I_{{\rm ang}}=4\pi\exp\bigl(C(\Delta)\bigr)$,
and the probability distribution of the spin-splitting modulus, after
incorporating $u=\Delta^{2}/(8k_{B}T\varepsilon_{p})$ and $\lambda_{p}$
to $C(\Delta)$, acquires the compact closed form given in the main
text (Eq.~(\ref{eq:Pparamag})). In the strict Gaussian limit $\lambda=0$
the quartic correction disappears and we recover the original DS probability
distribution of Eq.~(3.25) of Ref.~\cite{Dietl1983PRB}.

\section{Numerical evaluation of the partition function for finite $M_{0}\protect\neq0$}

For a ferromagnetic semiconductor such as GdN near its Curie temperature
$T_{c}\approx55\,\text{K}$, the spontaneous magnetization $M_{0}(T)$
is nonzero (below $T_{c}$) or can be made finite by a small external
field. In the fully resummed local theory ($\xi\to0$) the probability
distribution $P(\boldsymbol{\Delta})$ is still given by 
\begin{equation}
P(\boldsymbol{\Delta})=\mathcal{N}\exp\!\left(-\frac{\mathcal{H}_{1}}{k_{B}T}\right)\cosh\!\left(s\frac{\Delta}{k_{B}T}\right)Z[\boldsymbol{\Delta}],
\end{equation}
with the auxiliary functional $Z[\boldsymbol{\Delta}]$ resummed to
the exponential form~Eq.~(\ref{eq:Z-resum-local-tilded}) of the
main text. After the change of variables to modulus $\Delta$ and
angle $\theta$ (see main text), the angular integration yields the
closed-form result $I_{{\rm ang}}(\Delta;\Delta_{0})$ expressed with
$\operatorname{erf}$ or $\operatorname{erfi}$ (Appendix~B).

The normalization constant is then obtained from the remaining one-dimensional
radial integral 
\begin{equation}
Z=\int_{0}^{\infty}\Delta^{2}\cosh\left(\frac{s\Delta}{k_{B}T}\right)I_{{\rm ang}}(\Delta;\Delta_{0},\Phi)\,d\Delta,\label{eq:Z-radial-general}
\end{equation}
where the coefficients $A(\Delta)$, $B(\Delta)$, $C(\Delta)$ entering
$I_{{\rm ang}}$ are given explicitly by Eqs.~(\ref{eq:A})--(\ref{eq:C})
of Appendix~B and contain the full dependence on $M_{0}$ (via $\Delta_{0}$)
and on the quartic anharmonicity parameter $\lambda_{p}$.

Equation~(\ref{eq:Z-radial-general}) does not admit any elementary
closed-form solution for $M_{0}\neq0$. It is, however, suitable for
numerical quadrature. The integrand decays exponentially for large
$\Delta$ (dominated by the Gaussian and quartic pieces in $C(\Delta)$)
and is smooth for all physical parameters.

Once $Z$ is known, the BMP free-energy shift follows immediately:
\begin{equation}
\Delta F[\varphi]=\langle\varphi|\mathcal{H}_{1}|\varphi\rangle-k_{B}T\ln Z[\Phi[\varphi]],
\end{equation}
and the renormalized Schrödinger equation for the envelope $\varphi(\mathbf{r})$
is obtained by functional variation (Eq.~(\ref{eq:Sch-resummed})
of the main text). The most-probable spin splitting $\overline{\Delta}$
inside the polaron is found by numerically maximizing $\ln P(\Delta)$.

\section{Explicit non-local convolutions for the hydrogenic 1s envelope function}

For the hydrogenic 1s donor envelope wave function (normalized to
unity) 
\begin{equation}
\varphi(\mathbf{r})=\frac{1}{\sqrt{\pi a_{B}^{3}}}\exp\left(-\frac{r}{a_{B}}\right),\label{eq:phi-1s}
\end{equation}
the probability density is 
\begin{equation}
\rho(\mathbf{r})=|\varphi(\mathbf{r})|^{2}=\frac{1}{\pi a_{B}^{3}}\exp(-\alpha_{h}r),\label{eq:rho-1s}
\end{equation}
where we introduced the hydrogenic decay constant $\alpha_{h}\equiv2/a_{B}$
(unrelated to the exchange coupling constant $\alpha$ of Eq.~(\ref{eq:Delta0})
in the main text).

Its three-dimensional Fourier transform (with the convention $\tilde{f}(\mathbf{k})=\int f(\mathbf{r})\,e^{-i\mathbf{k}\cdot\mathbf{r}}\,d^{3}r$)
reads 
\begin{equation}
\tilde{\rho}(k)=\frac{\alpha_{h}^{4}}{(\alpha_{h}^{2}+k^{2})^{2}}.\label{eq:rho-tilde}
\end{equation}

The non-local auxiliary function appearing in the first-order anharmonic
correction (and in the definitions of $\mathcal{F}_{i}[\mathbf{J}=0]$
and $\mathcal{F}'_{i}[\mathbf{J}=0]$) is the convolution 
\begin{equation}
\Psi_{i}(\mathbf{r})=\int Q_{i}(\mathbf{r},\mathbf{r}')\rho(\mathbf{r}')\,d^{3}r',\label{eq:Psi-def}
\end{equation}
where the Yukawa propagator of the quadratic GLW theory is 
\begin{equation}
Q_{i}(\mathbf{r},\mathbf{r}')=\frac{\exp(-\kappa_{i}|\mathbf{r}-\mathbf{r}'|)}{4\pi\xi^{2}|\mathbf{r}-\mathbf{r}'|},\qquad\kappa_{i}=\frac{\sqrt{\bar{\mu}_{i}}}{\xi}.\label{eq:Q-i}
\end{equation}

In momentum space this convolution becomes a simple product. Performing
the inverse Fourier transform for the spherically symmetric case yields
the explicit one-dimensional integral representation 
\begin{equation}
\Psi_{i}(r)=\frac{\alpha_{h}^{4}}{2\pi^{2}\xi^{2}r}\int_{0}^{\infty}\frac{k\sin(kr)}{(k^{2}+\kappa_{i}^{2})(\alpha_{h}^{2}+k^{2})^{2}}\,dk.\label{eq:Psi-integral}
\end{equation}

The integral above admits a fully analytical closed-form evaluation
by the residue theorem (closing the contour in the upper half-plane
and picking up the simple pole at $k=i\kappa_{i}$ and the double
pole at $k=i\alpha_{h}$). For $\kappa_{i}\neq\alpha_{h}$, partial-fraction
decomposition of the rational function in $x=k^{2}$ 
\begin{equation}
\frac{1}{(x+\kappa_{i}^{2})(x+\alpha_{h}^{2})^{2}}=\frac{A}{x+\kappa_{i}^{2}}+\frac{B}{x+\alpha_{h}^{2}}+\frac{C}{(x+\alpha_{h}^{2})^{2}}
\end{equation}
with coefficients 
\begin{equation}
A=\frac{1}{(\kappa_{i}^{2}-\alpha_{h}^{2})^{2}},\qquad B=-\frac{1}{(\kappa_{i}^{2}-\alpha_{h}^{2})^{2}},\qquad C=\frac{1}{\kappa_{i}^{2}-\alpha_{h}^{2}}
\end{equation}
together with the known sine-transform integrals 
\begin{equation}
\int_{0}^{\infty}\frac{k\sin(kr)}{k^{2}+p^{2}}\,dk=\frac{\pi}{2}e^{-pr},\qquad\int_{0}^{\infty}\frac{k\sin(kr)}{(k^{2}+p^{2})^{2}}\,dk=\frac{\pi r}{4p}e^{-pr}
\end{equation}
immediately yields the explicit closed-form expression 
\begin{equation}
\Psi_{i}(r)=\frac{\alpha_{h}^{4}}{4\pi\xi^{2}r}\left[\frac{e^{-\kappa_{i}r}-e^{-\alpha_{h}r}}{(\kappa_{i}^{2}-\alpha_{h}^{2})^{2}}+\frac{r\,e^{-\alpha_{h}r}}{2\alpha_{h}(\kappa_{i}^{2}-\alpha_{h}^{2})}\right].\label{eq:Psi-closed}
\end{equation}
The special case $\kappa_{i}=\alpha_{h}$ is obtained by taking the
appropriate limit $\kappa_{i}\to\alpha_{h}$ of Eq.~(\ref{eq:Psi-closed}),
which yields the compact closed-form expression 
\begin{equation}
\Psi_{i}(r)=\frac{\alpha_{h}(1+\alpha_{h}r)\,e^{-\alpha_{h}r}}{32\pi\xi^{2}}.
\end{equation}

The Gaussian fluctuation parameter $\Phi_{i}$ (appearing in $\mathcal{F}_{i}$
and $\mathcal{F}'_{i}$) is likewise expressible in closed momentum-space
form: 
\begin{equation}
\Phi_{i}=\left(\frac{\alpha}{g\mu_{B}}\right)^{2}\frac{1}{2}\int\frac{d^{3}k}{(2\pi)^{3}}\tilde{\rho}(k)^{2}Q_{i}(k),\label{eq:Phi-nonlocal}
\end{equation}
where the extra factor $1/2$ comes from the definition of the diamond
convolution operator (Eq.~(\ref{eq:diamond}) of the main text) and
$Q_{i}(k)=1/(\bar{\mu}_{i}+\xi^{2}k^{2})$. This integral is also
evaluable analytically by the same residue technique (or numerically).
In the local limit $\xi\to0$ one recovers the standard DS expression
$\Phi=2k_{B}T\varepsilon_{p}$.

With the explicit $\Psi_{i}(r)$ at hand, all spatial integrals appearing
in the first-order non-local correction to $Z[\mathbf{J}=0]$ (Eqs.~(\ref{eq:Z-first-order-nonlocal-tilded})
and (\ref{eq:P-first-order-final-tilded}) of the main text) become
ordinary radial integrals of the form 
\begin{equation}
\int d^{3}z\,f\bigl(\Psi_{\parallel}(z),\Psi_{\perp}(z)\bigr)=4\pi\int_{0}^{\infty}r^{2}\,dr\,f\bigl(\Psi_{\parallel}(r),\Psi_{\perp}(r)\bigr),\label{eq:radial-int}
\end{equation}
where $f$ denotes the appropriate polynomial combination of $\mathcal{F}_{i}$
and $\mathcal{F}'_{i}$ (cubic and quartic terms). These integrals
are fully analytical (combinations of exponentials and polynomials
in $r$) for any finite correlation length $\xi$. The resulting effective
(non-local) generalizations of $\overline{I}_{3}$ and $\overline{I}_{4}$
entering the first-order correction to $P(\boldsymbol{\Delta})$ are
therefore known in closed form for the realistic 1s envelope.

In the strict local limit $\xi\to0$ ($\kappa_{i}\to\infty$) the
expression~(\ref{eq:Psi-closed}) reduces (after the ultraviolet
regularization $q_{i}=0$) to the local result used in the resummed
theory of the main text. When additionally $\lambda\to0$, the whole
non-local first-order theory reduces exactly to the original Dietl--Spałek
Gaussian theory (both for arbitrary envelope functions and specifically
for the 1s hydrogenic function).

This appendix completes the analytical toolkit needed for a fully
quantitative, non-local treatment of non-Gaussian fluctuations with
finite spin correlation length while preserving exact reduction to
the original Dietl--Spałek Gaussian theory when $\lambda=0$.

 \bibliographystyle{apsrev4-1}
\bibliography{HB_FBMP}

\end{document}